\documentclass[a4paper,11pt]{article}
\pdfoutput=1
\usepackage{jcappub}
\usepackage{caption}
\usepackage{subcaption}
\usepackage{graphicx}
\usepackage{hyperref}
\usepackage{amsmath}
\usepackage{mathtools}
\usepackage{ulem}
\usepackage{upgreek}
\usepackage{tablefootnote}
\usepackage{soul}
\usepackage{color}
\usepackage{verbatim}
\usepackage{float}
\usepackage{bm}
\usepackage{placeins}
\usepackage{enumitem}
\usepackage[export]{adjustbox}

\newcommand{\bk}{\mathbf{k}}

\begin{document}

\title{Chern-Simons gravitational term coupled to a spectator field}

\author[a]{Giorgio Orlando}

\affiliation[a]{Faculty of Physics, Astronomy and Applied Computer Science, Jagiellonian University, 30-348 Krakow, Poland}
\emailAdd{giorgio.orlando@uj.edu.pl}

\abstract{The Chern-Simons gravitational term is typically coupled to the inflaton field during inflation. In this work, we explore an alternative scenario where the Chern-Simons term is minimally coupled to a massive spectator field, $\sigma$, within a multi-field inflationary framework where the spectator field is linearly coupled to the inflaton. We derive the corresponding parity-violating cubic interaction Lagrangian involving the curvature, spectator and tensor perturbations and compute how it affects primordial scalar-tensor bispectra. We find that these correlators yield distinctive parity-odd shapes. Perturbativity and consistency bounds are derived, constraining the couplings and the amplitude of the associated non-Gaussianities.} 

\maketitle

\section{Introduction}

The inflationary epoch offers a unique laboratory for studying fundamental interactions at energy scales far beyond the reach of terrestrial colliders. The simplest models of inflation—namely, single-field slow-roll models—feature a single scalar field evolving under Einstein gravity \cite{Brout:1977ix,Guth:1980zm,Mukhanov:1981xt,Linde:1981mu}. These models predict the generation of primordial scalar and tensor perturbations from quantum fluctuations in the early universe, which set the initial conditions for the subsequent cosmic evolution. The most robust probes of these perturbations are the anisotropies in the Cosmic Microwave Background (CMB) \cite{Zaldarriaga:1996xe,Hu:1997hp}. The Planck experiment currently provides the most stringent constraints on inflationary models \cite{Planck:2018jri,Planck:2019kim,Campeti:2022vom}.

Observational data are consistent with a nearly Gaussian distribution of primordial perturbations. While simple inflationary models align well with this paradigm, more complex scenarios involving additional fields or modifications to Einstein gravity are also viable. These extended models typically produce non-Gaussian signatures (see, e.g., the review \cite{Celoria:2018euj}), which could help distinguish among competing theories if future observations detect primordial non-Gaussianity. Moreover, upcoming CMB experiments aim to detect tensor perturbations via the so-called $B$-modes. Measuring the amplitude of tensor perturbations is crucial to identifying the correct inflationary model and offers a path to detecting scalar-tensor non-Gaussianities \cite{Bartolo:2018elp,Shiraishi:2019yux,Philcox:2023xxk}, which are currently either weakly constrained or unconstrained. The latest bound on the tensor-to-scalar ratio, $r<0.037$ at $95\%$ confidence level, comes from a combination of Planck and BICEP/Keck Array data \cite{Campeti:2022vom}. Future experiments aiming to improve these measurements include LiteBIRD \cite{LiteBIRD:2022cnt}, CMB-S4 \cite{Abazajian:2019eic}, the Simons Observatory \cite{SimonsObservatory:2018koc}, and the BICEP Array \cite{Moncelsi:2020ppj}.

In this work, we study the impact of the gravitational Chern-Simons (gCS) term coupled to a scalar field during inflation (see, e.g., \cite{Lue:1998mq,Alexander:2004wk,Satoh:2010ep,Dyda:2012rj,Bartolo:2017szm,Creque-Sarbinowski:2023wmb,Christodoulidis:2024ric} for a previous literature on the topic). This term is often coupled minimally to the inflaton field. While it does not alter the classical background evolution, it can induce parity-violating signatures in cosmological correlators involving tensor perturbations and in the scalar trispectrum. A well-known issue with this setup is the emergence of ghost-like tensor modes above a characteristic scale—the so-called Chern-Simons scale $M_{CS}$—due to the sign flip in the quadratic kinetic term for one helicity state \cite{Dyda:2012rj}. To avoid ghost instabilities during inflation, one must impose $H/M_{CS} \ll 1$, which significantly suppresses observable signatures in the tensor power spectrum and higher-order correlation functions as well.

Motivated by this limitation, here we consider a two-field slow-roll framework in which the gCS term is linearly coupled to a massive spectator field $\sigma$. Also, we introduce a kinetic coupling between the spectator field and the inflaton field which respects the approximate shift symmetry of the inflaton. This yields a linear coupling between the two fields, enabling the transfer of features from the $\delta\sigma$ correlators to those involving the curvature perturbation $\zeta$. Moreover, we assume that the resulting centrifugal force in the spectator field background equation of motion generated by the kinetic coupling is compensated by the spectator field self-potential. It results a quasi-static motion for $\sigma_0$, and we derive appropriate bounds on the strength of the kinetic coupling to ensure that the spectator field contributes negligibly to either the background dynamics and the scalar power spectrum.

In this setup, the gCS term coupled to the spectator field can provide parity-violating signatures in both the quadratic and higher-order Lagrangian: however, we demonstrate that while quadratic corrections to the tensor sector are governed by the product of the time derivative of the spectator field $\dot\sigma_0$ and the gCS-$\sigma$ coupling constant $\kappa_{CS}$, i.e. $\kappa_{CS}\dot\sigma_0$, the leading cubic interactions depend solely on $\kappa_{CS}$. This motivates a detailed exploration of these interactions in comparison with the standard inflaton-gCS scenario.

Perturbations of additional scalar fields, relative to those of the inflaton, can generate significant CMB isocurvature perturbations, even if their background dynamics does not substantially modify the inflationary evolution. However, CMB observations require isocurvature perturbations to be much smaller than curvature perturbations. To ensure this, we assume that the spectator field decays before the end of inflation. This is achievable if the mass of $\sigma$, determined by the second order derivative of its potential $V(\sigma)$, is sufficiently large, for example of the order of the Hubble parameter, $m_{\sigma} \sim H$. The latter is the more natural scenario unless inflation lasts exceptionally long.

We show that coupling the gCS term to $\delta\sigma$ introduces novel parity-violating interactions of the forms $\delta\sigma\delta\sigma h$,  $\zeta \delta\sigma h$ and $ \delta\sigma h  h$, where $h$ denotes tensor perturbations. Combined with the linear $\delta\sigma\zeta$ coupling, these interactions can generate parity-odd bispectra $\langle \zeta\zeta h\rangle$ and $\langle \zeta hh\rangle$. We compute these bispectra using Schwinger-Keldysh (SK) diagrammatic rules~\cite{Weinberg:2005vy,Chen:2017ryl}, first deriving in-in expressions in terms of the generic $\delta\sigma$ mode functions, and then evaluating the resulting time integrals analytically for arbitrary $\nu$ by virtue of recently derived factorization theorems on parity-odd correlators~(see e.g. \cite{Stefanyszyn:2023qov,Thavanesan:2025kyc}). 

We find that non-zero $\langle \zeta\zeta h\rangle$ and $\langle \zeta hh\rangle$ parity-odd bispectra emerge when the spectator field has a nonzero mass. The amplitudes of these bispectra depend on both the 
$\delta\sigma$-gCS and $\delta\sigma \zeta$ coupling constants. Imposing perturbativity and the ghost-avoidance constraint we delineate the region of parameter space in which parity-odd non-Gaussian signatures are allowed and quantify how these theoretical bounds restrict the amplitude of the induced bispectra.

The paper is organized as follows. Section~\ref{sec:model} introduces the theoretical framework, i.e. a two-field inflationary model with a slow-rolling inflaton, and a spectator field kinetically coupled to the inflaton and linearly coupled to the gravitational Chern-Simons term. Here, we analyze implications for the background and the perturbative expansion up to second order. 
Section~\ref{sec:cubic_comp} derives the cubic interaction Lagrangian produced by the gravitational Chern-Simons term coupled to the spectator field. In Section~\ref{sec:Feynman}, we compute the resulting scalar-tensor bispectra and estimate bounds on the coupling constants from perturbativity and consistency of the theory. Finally, Section~\ref{sec:conclusion} summarizes our findings.

\section{The model}
\label{sec:model}
\subsection{Slow-roll inflation with a spectator field}
\label{sec:model_2fields}
Here we introduce an inflationary theory with two scalar fields, one of which represents the usual inflaton field, $\phi$, which relates to the gauge-invariant curvature perturbation $\zeta$, and the other a spectator field, $\sigma$. The inflaton field follows the normal slow-roll dynamics under a potential $V(\phi)$, while we introduce an independent potential $V(\sigma)$ for $\sigma$. Moreover, we introduce a minimal mixing operator between the inflaton and spectator field that respects the approximate shift symmetry of the inflaton field. This is achieved by adding a correction to the inflaton kinetic term linear in $\sigma$. The action of the model reads
\begin{align} \label{eq:quasi-slow}
S_{\rm kin} = \int d^4 x \sqrt{-g} \, \left[ \frac{M^2_{pl}}{2} R^{(4)} - \frac{1}{2}\Big(1+\kappa_\sigma\sigma\Big) (\partial_\mu \phi)(\partial^\mu \phi)  - \frac{1}{2}(\partial_\mu \sigma)(\partial^\mu \sigma) - V(\phi) - V(\sigma) \right] \, , 
\end{align}
where $M_{pl}$ is the reduced Planck Mass, $R^{(4)}$ is the scalar curvature, and the corrective operator introduced to the kinetic term of the inflaton provides a coupling between $\phi$ and $\sigma$, with $\kappa_\sigma>0$ a coupling constant\footnote{With this choice of sign for the coupling constant, we need $\sigma_0\ge0$ to ensure that the inflaton kinetic term always has the right sign.}. The limit $\kappa_\sigma\rightarrow 0$ corresponds to the decoupling limit. Here, the signature of the metric is $(-1,1,1,1)$.

In this setup the background e.o.m. of $\phi_0$ and $\sigma_0$ read
\begin{align} \label{eq:phi_eom}
(1+\kappa_\sigma \sigma_0)\ddot \phi_0 + \left[3 H (1+\kappa_\sigma \sigma_0) + \kappa_\sigma \dot\sigma_0\right]  \dot \phi_0 + \partial_\phi V(\phi_0)  = 0\, , 
\end{align}
and 
\begin{align} \label{eq:sigma_eom}
\ddot \sigma_0 + 3 H  \dot \sigma_0 + \partial_\sigma V(\sigma_0) = \frac{1}{2} \kappa_\sigma \dot\phi_0^2\, , 
\end{align}
where $H = \dot a/a$, with $a$ the Robertson-Walker scale factor. The dot here means differentiation with respect to the cosmological time $t$.

Also, the Friedmann equations read
\begin{align}
3 M^2_{pl} H^2 &= \frac{\dot \phi_0^2}{2}(1+\kappa_\sigma \sigma_0) +\frac{\dot \sigma_0^2}{2} + V(\phi_0) + V(\sigma_0) \, , \\
M^2_{pl} \dot H &= -\frac{\dot \phi_0^2}{2}(1+\kappa_\sigma \sigma_0) -\frac{\dot \sigma_0^2}{2}  \, . 
\end{align}
Here we want $\sigma$ to be a spectator field, therefore its contribution towards the energy density and pressure of the universe are supposed to be much subdominant to those of the inflaton field. In other words, the relations
\begin{align}
\kappa_\sigma \sigma_0 &\ll 1\, , \label{eq:Fried1} \\ 
V(\sigma_0) &\ll V(\phi_0) \, , \label{eq:Fried2}\\
|\dot\sigma_0| &\ll |\dot\phi_0|  \label{eq:Fried3}\, , 
\end{align}
must be realized during inflation.

Furthermore, in order for the field $\sigma$ not to modify significantly the inflaton e.o.m., we need to assume 
\begin{align}
|\kappa_\sigma \dot\sigma_0| \ll 3H  \label{eq:phiback}\, , 
\end{align}
as well. 

If $\kappa_\sigma$ is large enough, in principle the centrifugal force on the r.h.s. of Eq. \eqref{eq:sigma_eom} is capable to provide a significant dynamics to the field $\sigma_0$ which in turn might spoil the inflationary background. In order to compensate for the effects of this force the potential $V(\sigma)$ is fixed so that at the beginning of inflation 
\begin{align} \label{eq:centr_compensated_V}
\partial_\sigma V(\sigma_0) = \frac{1}{2} \kappa_\sigma \dot\phi_0^2  \, . 
\end{align}
In this work we will assume a simple form for the $\sigma$-potential:
\begin{align} \label{eq:sigmapot}
V(\sigma) = \frac{m_\sigma^2 \,\sigma^2}{2} \, ,
\end{align}
where $m_\sigma$ represents the mass of the spectator field. Therefore, with the in choice Eq. \eqref{eq:sigmapot}, Eq. \eqref{eq:centr_compensated_V} reads
\begin{align} \label{eq:parameters_satisf}
 m^2_\sigma \sigma_0  = \frac{1}{2} \kappa_\sigma \dot\phi_0^2  \, . 
\end{align}
This is the condition that ensures $\dot \sigma_0 = 0$ at the onset of inflation. Note however that during inflation $\dot \phi_0$ slowly  evolves in time, and therefore this provides a non-trivial slow-dynamics to $\sigma_0$ which adiabatically follows the change in $\dot \phi_0$. The quasi-static value of $\sigma_0$ in terms of the parameters of the theory is determined by Eq. \eqref{eq:parameters_satisf} as
\begin{align} \label{eq:sigma_0}
\sigma_0  = \frac{\kappa_\sigma \dot\phi_0^2}{2m^2_\sigma}  \, ,
\end{align}
and therefore its time derivative is as
\begin{align} \label{eq:sigma_0dot}
\dot\sigma_0  = \frac{\kappa_\sigma \dot\phi_0 \ddot\phi_0}{m^2_\sigma}  \, . 
\end{align}
Now, for consistency with the initial assumptions, let's verify the bounds on $\kappa_{\sigma}$ that we need to impose so that Eqs. \eqref{eq:Fried1}-\eqref{eq:phiback} are satisfied:
\begin{itemize}
\item condition \eqref{eq:Fried1}: 
by plugging Eq. \eqref{eq:sigma_0} into condition \eqref{eq:Fried1} we get 
\begin{align} 
\frac{1}{2} \frac{\kappa_\sigma^2 \dot\phi_0^2}{m^2_\sigma} \ll 1 \, \, ,
\end{align}
which translates into the following bound to $\kappa_\sigma$
\begin{align}  \label{eq:condition1}
\kappa_\sigma  \ll \sqrt{2} \frac{m_\sigma}{|\dot\phi_0|} \, .
\end{align}
\item condition \eqref{eq:Fried2}: by plugging Eq. \eqref{eq:sigma_0} into condition \eqref{eq:Fried2} we get
\begin{align}  
\frac{1}{8} \frac{\kappa^2_\sigma \dot\phi_0^4}{m^2_\sigma}  \ll V(\phi_0) \, .
\end{align}
Note that when the previous Eq. \eqref{eq:condition1} is realized, then the left-hand side of the previous equation is bounded as 
\begin{align}  
\frac{1}{8} \frac{\kappa^2_\sigma \dot\phi_0^4}{m^2_\sigma}  \ll \frac{1}{4} \dot\phi_0^2 \, .
\end{align}
Since 
\begin{align}  
\frac{1}{4} \dot\phi_0^2 \ll V(\phi_0) 
\end{align}
is consistent with the inflaton slow-roll dynamics, it follows that when Eq. \eqref{eq:Fried1} is realized, also \eqref{eq:Fried2} is realized automatically. 
\item condition \eqref{eq:Fried3}: by plugging Eq. \eqref{eq:sigma_0dot} into condition \eqref{eq:Fried3} we get 
\begin{align}  
\left|\frac{\kappa_\sigma \dot\phi_0 \ddot\phi_0}{m^2_\sigma}\right| \ll |\dot\phi_0| \, .
\end{align}
This provides the following alternative bound on $\kappa_\sigma$
\begin{align}  
\kappa_\sigma \ll \frac{m^2_\sigma}{|\ddot\phi_0|} \, .
\end{align}
\item condition \eqref{eq:phiback}: by plugging Eq. \eqref{eq:sigma_0dot} into condition \eqref{eq:phiback} we get
\begin{align}  
\left|\frac{\kappa^2_\sigma \dot\phi_0 \ddot\phi_0}{m^2_\sigma}\right| \ll 3 H\, .
\end{align}
Therefore, condition \eqref{eq:phiback} requires 
\begin{align}  
\kappa_\sigma \ll \sqrt{\frac{3 H}{|\dot\phi_0 \ddot\phi_0|}} m_\sigma \, ,
\end{align}
which is weaker than the bound in Eq. \eqref{eq:condition1} due to the inflaton slow-roll dynamics. 
\end{itemize}
Therefore, we have shown that the spectator field provides a negligible contribution to the inflation dynamics providing that $\kappa_\sigma$ simultaneously respect the following two boundaries
\begin{align}  
\kappa_\sigma  \ll \begin{cases} \sqrt{2} \frac{m_\sigma}{|\dot\phi_0|}  \, , \\
 \frac{m^2_\sigma}{|\ddot\phi_0|}  \, . \label{eq:constraint_kappa}
\end{cases}
\end{align}
It is helpful to express the bounds in Eq. \eqref{eq:constraint_kappa} in terms of slow-roll parameters
\begin{align} \label{eq:epsilon}
\epsilon &= \frac{1}{2} \left(\frac{M_{pl} \, \partial_\phi V}{V} \right)^2 \simeq \frac{1}{2} \frac{\dot \phi_0^2}{H^2} \frac{1}{M^2_{pl}} \, , \\
\eta &= \frac{M^2_{pl} \, \partial_{\phi\phi}V}{V} \simeq - \frac{\ddot \phi_0}{\dot \phi_0 H} + \frac{1}{2} \frac{\dot \phi_0^2}{H^2} \frac{1}{M^2_{pl}} \label{eq:eta}\, . 
\end{align}
From Eq. \eqref{eq:eta} we get
\begin{align}
\ddot \phi_0  \simeq  \left(\epsilon - \eta\right) \, \sqrt{2 \epsilon} \, M_{pl} H^2 \, . 
\end{align}
It follows that in terms of slow-roll parameters Eq. \eqref{eq:constraint_kappa} reads
\begin{align}  
\kappa_\sigma  \ll \begin{cases} \frac{m_\sigma}{H}\frac{\epsilon^{-1/2}}{M_{pl}} \, , \\
 \frac{m^2_\sigma}{H^2}\frac{(2\epsilon)^{-1/2} |\epsilon-\eta|^{-1/2}}{M_{pl}} \, . \label{eq:constraint_kappa2}
\end{cases}
\end{align}
From the previous equation, it is evident that $\kappa_\sigma$ is constrained the most when spectators have a mass much smaller than the Hubble parameter ($m_\sigma \ll H$). On the other hand, when $m_\sigma \sim H$ the first constraint in Eq. \eqref{eq:constraint_kappa2} is stronger than the second one.


We conclude this investigation of the background dynamics by pointing out that one could consider potentials $V(\sigma)$ different than our choice in Eq. \eqref{eq:sigmapot} resulting possibly in a more relaxed bound on $\kappa_\sigma$ with respect to Eq. \eqref{eq:constraint_kappa2}. In this paper we are not interested to explore more complicated potentials. Also, as we will see below, the constraint in Eq. \eqref{eq:constraint_kappa2} is enforced by perturbativity considerations, even if a more relaxed constraint is introduced by modifying $V(\sigma)$.

Now, we switch our focus to the perturbations sector of our model. When doing perturbation theory it is convenient to re-write the action \eqref{eq:quasi-slow} using the metric in the ADM form~\cite{Arnowitt:1962hi} 
\begin{align} \label{eq:ADM}
ds^2 &= g_{\mu\nu} dx^\mu dx^\nu = -(N^2 - N^i N_i \,) \, dt^2 + 2 N_i \, dx^i dt + g_{ij} \, dx^i dx^j \, , \nonumber \\
 &= g^{\mu\nu} dx_\mu dx_\nu = - N^{-2} \, dt^2 + 2  N^i N^{-2}  \, dx_i d t + \left(g^{ij} - \frac{N^i N^j}{N^2}\right) \, dx_i dx_j \, . 
\end{align}
Employing this metric the action reads 
\begin{align} \label{eq:quasi-slow2}
S_{\rm kin} = \frac{1}{2}\int d^4 x \sqrt{-g^{(3)}} \, N \Big[ & M^2_{pl} R^{(3)} + N^{-2}M^2_{pl}\left(E_{ij} E^{ij} - E^2\right) - \Big(1+\kappa_\sigma\sigma\Big) (\partial_\mu \phi)(\partial^\mu \phi)  \nonumber  \\
& \qquad\qquad - (\partial_\mu \sigma)(\partial^\mu \sigma) - 2V(\phi) - 2V(\sigma) \Big] \, , 
\end{align}
where as usual $R^{(3)}$ is the scalar curvature computed with the three-dimensional metric $g_{ij}$ and
\begin{align}
E_{ij} &= \frac{1}{2} \left( \dot h_{ij} - \nabla_i N_j - \nabla_j N_i \right) \, , \\
E &= E_{ij} \, g^{ij} \, . 
\end{align}
From now on we fix the uniform inflaton (or comoving) gauge (see e.g. \cite{Maldacena:2002vr})
\begin{align} \label{eq:comoving_gauge}
\phi = \phi_0  \, , \qquad  \sigma = \sigma_0 + \delta \sigma \, , \qquad g_{ij} &= a^2 e^{2 \zeta} \exp[h]_{ij} \, ,  \nonumber \\
e^{2 \zeta} &= 1+ 2 \zeta + 2 \zeta^2+... \, , \nonumber \\
\exp[h]_{ij} &= \delta_{ij}+h_{ij}+\frac{1}{2} h_{in} \,h^n_{j}+... \, ,
\end{align}
where $h_{ij}$ indicates the usual transverse and traceless tensor perturbations.
In this gauge at leading order in slow-roll the first order solutions for the constraint fields $N$ and $N_i$ read
\begin{align} \label{eq:constraints_quasi}
N &= 1+\frac{\dot \zeta}{H} \, , \nonumber \\
N_i &= \partial_i \psi \, , \qquad \psi = - \frac{\zeta}{H} + \chi \, , \qquad \partial^2\chi = a^2 \epsilon \, \left(\dot \zeta - \frac{1}{2} H \kappa_\sigma \,\delta \sigma\right) \, ,
\end{align}
where we have accounted for the new degree of freedom introduced by $\delta\sigma$ (see e.g. \cite{Chen:2009zp} for an analogous computation). Expanding the action \eqref{eq:quasi-slow2} in perturbations in this gauge, the free quadratic Lagrangian for the fields $\zeta$ and $\delta\sigma$ reads 
\begin{align}
\mathcal L^{\zeta}_2 &= M^2_{pl} \, \epsilon \left[a^3 \dot \zeta^2 - a (\partial \zeta)^2\right] \, , \\ 
\mathcal L^{\sigma}_2 &= \frac{a^3}{2} \dot{\delta\sigma}^2 - \frac{a}{2} (\partial_i\delta\sigma)^2 - \frac{a^3}{2} m_{\sigma}^2  \delta\sigma^2 \label{eq:quadratic_sigma}  \, . 
\end{align}
This is the action for the massless curvature perturbation field $\zeta$ and a massive perturbation field $\delta\sigma$ with mass $m_{\sigma}^2 = \partial^2_\sigma V(\sigma_0)$.  
As usual, after switching to the conformal time and going in Fourier space, we canonically quantize the fields as 
\begin{align} 
\zeta(\tau, \mathbf x) =  \int \frac{d^3 \mathbf k}{(2 \pi)^3} \, \zeta(\tau,{\mathbf k})  \, e^{i \mathbf{k} \cdot \mathbf{x}} \, , \qquad \zeta(\tau, \mathbf k) = \, \left[ u_{\zeta}(\tau,{\mathbf k}) \, \hat a_\lambda(\mathbf k) + u^*_{\zeta} (\tau,-{\mathbf k}) \, \hat a^{\dagger}_\lambda(-\mathbf k) \right] \, , \nonumber \\
\delta\sigma(\tau, \mathbf x) =  \int \frac{d^3 \mathbf k}{(2 \pi)^3} \, \delta\sigma(\tau,{\mathbf k})  \, e^{i \mathbf{k} \cdot \mathbf{x}} \, , \qquad \delta\sigma(\tau, \mathbf k) = \, \left[ v_{\sigma}(\tau,{\mathbf k}) \, \hat {\tilde a}_\lambda(\mathbf k) + v^*_{\sigma} (\tau,-{\mathbf k}) \, \hat {\tilde a}^{\dagger}_\lambda(-\mathbf k) \right] \, , 
\end{align}
where $\tau$ denotes the conformal time. At leading order in slow-roll the free-solutions for the mode functions with Bunch-Davies vacuum initial conditions are \cite{Maldacena:2002vr}
\begin{align} \label{eq:uzeta}
u_{\zeta}(\tau,{\mathbf k}) =& \frac{H}{M_{pl}\sqrt{4 \epsilon \, k^3}} (1+i k \tau) e^{-ik\tau} \, ,
\end{align}
and \cite{Chen:2009zp} 
\begin{align} \label{eq:vsigma_m}
v_{\sigma}(\tau,{\mathbf k}) = - i e^{i (\nu +\frac{1}{2}) \frac{\pi}{2}} \frac{\sqrt \pi}{2}\, H (-\tau)^{3/2} \, H^{(1)}_\nu(-k\tau)  \qquad \mbox{when } \, m_\sigma^2/H^2 \leq 9/4 \, ,
\end{align}
where $\nu = \sqrt{9/4 - m_\sigma^2/H^2}$, and $H^{(1)}_\nu(-k\tau)$ is the Hankel function of the first kind. When $m_\sigma^2/H^2 > 9/4$ the contributions of the spectator field to the correlation functions involving the curvature field are exponentially suppressed by factors of $e^{- m_{\sigma}/H}$, via a mechanism analogous to the Boltzmann suppression. Therefore, here we will not consider this case. 

As most of the interesting signatures of the present model are generated after the horizon crossing of the spectator field $\delta\sigma$, it is interesting to look at the behavior of the solution in Eq. \eqref{eq:vsigma_m} in the super-horizon limit $-k\tau \rightarrow 0$. For $m_\sigma^2/H^2 \leq 9/4$ we have 
\begin{equation} \label{eq:asympthotic_sigma}
v^{\sigma}_{\mathbf k} \longrightarrow \begin{cases} 
- e^{i (\nu +\frac{1}{2}) \frac{\pi}{2}} \frac{2^{\nu-1}}{\sqrt\pi} \Gamma(\nu) \frac{H}{k^\nu} (-\tau)^{-\nu+3/2} \, , \qquad 0 < \nu \leq 3/2 \, ,\\
e^{i \frac{\pi}{4}} \frac{1}{\sqrt\pi} H (-\tau)^{3/2} \ln(-k\tau) \, , \qquad\qquad\qquad \nu = 0 \, . 
\end{cases}
\end{equation}
Notice the decay factor $(-\tau)^{3/2}$ in the $-k\tau \rightarrow 0$ limit. This leads to the decay of the $\delta\sigma$ perturbations once they reach super-horizon scales. The decay is faster the larger is the mass of $\delta\sigma$ (the smaller is $\nu$). In the massless limit ($\nu \rightarrow 3/2$), the decay disappears, and the spectator field perturbations survive throughout inflation. Before their complete decay or the end of inflation, the spectator field is capable of sourcing curvature perturbations. This is possible thanks to the linear coupling between $\delta\sigma$ and $\zeta$ which is generated by expanding in perturbations the action in Eq. \eqref{eq:quasi-slow2}. By doing so, at leading order in slow-roll we get
\begin{align} \label{eq:linear_coupling_orig}
\mathcal L^{\sigma\zeta}_2 &= - a^3 \frac{\dot \phi_0^2}{H} \kappa_\sigma \,\delta \sigma \dot \zeta  \, . 
\end{align}
For the purpose of the computations below it is useful to re-write this interaction after we switch to conformal time $\tau$ , use $a\simeq -1/(H \tau)$ and the definition of slow-roll parameter~\eqref{eq:epsilon}. We get
\begin{align} \label{eq:intst}
\mathcal L^{\sigma\zeta}_2 &= 2 \frac{M^2_{pl}}{H^2} \frac{\epsilon \kappa_\sigma}{\tau^3} \, \delta \sigma \zeta'  \, ,
\end{align}
where the prime $'$ indicates differentiation with respect to the conformal time. 

\begin{figure}
        \centering
        \includegraphics[width=.7\textwidth]{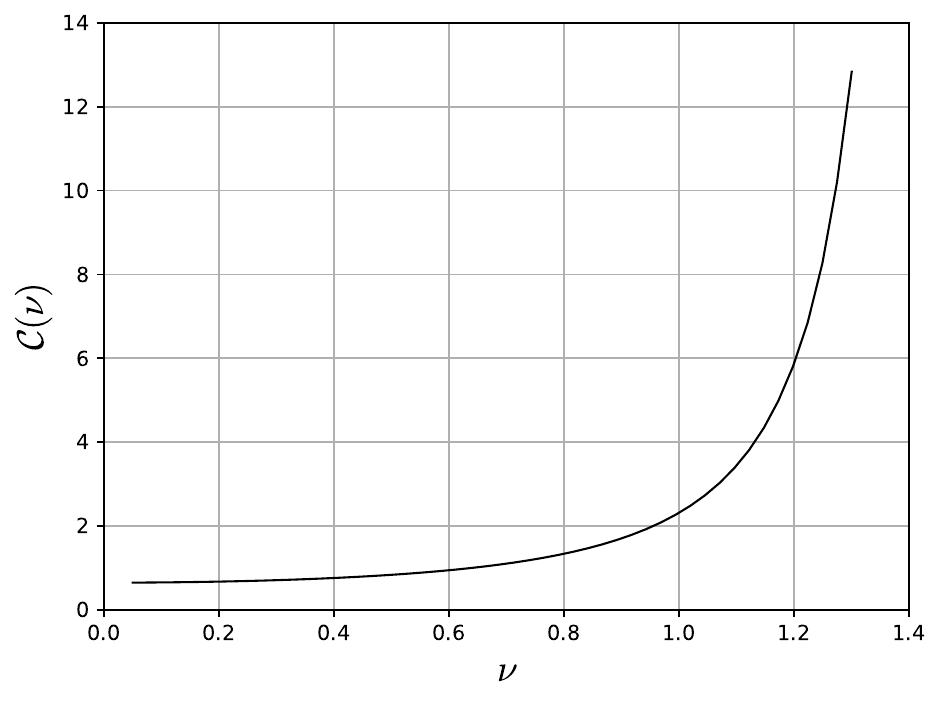} 
    \caption{Parameter $\mathcal{C}(\nu)$ as in Eq. \eqref{eq:Cnu}.}  \label{fig:Cnu}
\end{figure}

The interaction \eqref{eq:intst} induces a correction at tree-level to the scalar power spectrum. For the type of interaction in Eq. \eqref{eq:intst} this has been computed e.g. in Ref. \cite{Chen:2009zp}. As a result of this correction, if we believe in perturbation theory we shall impose~\cite{Chen:2009zp}\footnote{The exact numerical coefficient differs from Eq. (3.10) of \cite{Chen:2009zp} as here the coefficient of the interaction in Eq. \eqref{eq:intst} is different.}
\begin{align} \label{eq:radiative_control_mass_precise}
4 \mathcal C(\nu)\,\epsilon \,(\kappa_\sigma M_{pl})^2 \ll 1 
\end{align}
which is equivalent to
\begin{align} \label{eq:contraint_R}
\kappa_{\sigma} \ll \frac{\epsilon^{-1/2}}{2M_{pl} \, \sqrt{\mathcal C(\nu)}}  \, ,
\end{align}
where $\mathcal C(\nu)$ reads \cite{Chen:2009zp}
\begin{align} \label{eq:Cnu}
\mathcal C(\nu) =& \frac{\pi}{4} {\rm Re}\Big[\int_0^{\infty} dx_1 \int_{x_1}^{\infty} d x_2 \, \Big( 
x_1^{-1/2} H_\nu^{(1)}(x_1) e^{i x_1} x_2^{-1/2} H_\nu^{(2)}(x_2) e^{-i x_2} \nonumber \\
&\qquad\qquad\qquad\qquad\qquad\qquad -  x_1^{-1/2} H_\nu^{(1)}(x_1) e^{-i x_1} x_2^{-1/2} H_\nu^{(2)}(x_2) e^{-i x_2}\Big) \Big] \, .
\end{align}
This quantity is plotted as a function of $\nu$ in Fig. \ref{fig:Cnu}.
Note that when $m_{\sigma} \sim H$, condition \eqref{eq:contraint_R} is practically equivalent to \eqref{eq:constraint_kappa2}. When $m_{\sigma} \ll H$, $\sqrt{C(\nu)}$ tends to explode faster than  $(H/m_{\sigma})^2$. Therefore, in such case condition \eqref{eq:contraint_R} is stronger than \eqref{eq:constraint_kappa2}. However, here there is a subtlety, as a limit case arises when the spectator field is (almost) massless: if we take the value of $\mathcal C(\nu)$ literally, we would end up with a divergent result in the massless limit $\nu \rightarrow 3/2$, which is clearly nonphysical as it would provide a divergent correction to the power spectrum. Instead, in such a case, the quantity $\mathcal C(\nu)$ should be replaced by $N^2_k$, where $N_k$ denotes the number of e-foldings from the time the scale corresponding to $k$ leaves the horizon during inflation and the end of inflation. In other words, the maximum correction to the scalar power spectrum is restricted by the duration of inflation. In this peculiar case we need to apply the condition  
\begin{align} \label{eq:radiative_control}
4\epsilon \, (\kappa_\sigma M_{pl} N_k)^2 \ll 1 \, \quad \mbox{when } m_{\sigma} \simeq 0 \, .
\end{align}
This condition is important to prevent cosmological correlators to grow uncontrollably as the mass of the spectator field approaches 0. As a result, when $m_\sigma \rightarrow 0$, indeed condition \eqref{eq:constraint_kappa2} predominates, resulting in the necessity of taking $\kappa_{\sigma} \rightarrow0$ to prevent spoiling the inflation dynamics. Therefore, in our model (almost) massless spectators should be decoupled from the inflaton field and an alternative transfer mechanism should be studied. This suggest we should focus on the case $m_\sigma \sim H$, which is also motivated by CMB observations as explained in the introduction.

In general, if we work perturbatively and we assume little corrections to the scalar power spectrum, combining cubic $\sigma-\zeta$ interactions generated in the action \eqref{eq:quasi-slow2} with the linear coupling in Eq. \eqref{eq:intst} is unable to provide sizable non-Gaussianities \cite{Chen:2009zp}. On the contrary, combining the mediation in Eq. \eqref{eq:intst} with appropriate $\delta \sigma^3$ self-interactions coming from the Taylor expansion of the potential $V(\sigma)$ are capable to source sizable scalar non-Gaussianities at tree level when $\partial_\sigma^3 V(\sigma_0)$ is large enough \cite{Chen:2009zp}. The basic intuition underlying this result is that $\sigma$ self-interactions do not modify the $\zeta$ power spectrum at tree level, thereby allowing for a less restrictive perturbative bound. However, this effect is not the focus of the present work, and we refer the reader to the existing literature for further details. It is worth noting that one can always choose the potential $V(\sigma)$ such that this additional contribution to scalar non-Gaussianities is either absent or subdominant compared to that arising from $\zeta^3$ self-interactions, effectively reproducing the single-field inflation result. An explicit example is provided by our choice of the potential in Eq.~\eqref{eq:sigmapot}, for which $\partial_\sigma^3 V(\sigma_0) = 0$.

Moving to the tensor sector, at quadratic level this is untouched with respect to single-field slow-roll inflation. By expanding the action \eqref{eq:quasi-slow2} in tensor perturbations we get the usual Lagrangian (here already expressed in conformal time)
\begin{equation}
\mathcal L_2^h = \frac{M^2_{pl}}{8} \, a^2 \left[ (h_{ij})' (h^{ij})' + (\partial^2 h_{ij}) (h^{ij}) \right] \, , 
\end{equation}
where Latin contractions are made with the Kronecker-delta. 

In Fourier space and after the canonical quantization the tensor perturbations read
\begin{equation}
h_{ij}(\tau, \mathbf x) = \sum_{\lambda = R/L} \frac{d^3 \mathbf k}{(2 \pi)^3} \, \left[ h_\lambda (\tau,{\mathbf k}) \, e^{\lambda}_{ij}(\mathbf k)  \right] \, e^{i \mathbf{k} \cdot \mathbf{x}} \, , \quad h_\lambda(\tau, \mathbf k) = \, \left[ u_\lambda (\tau,{\mathbf k}) \, \hat b_\lambda(\mathbf k) + u^*_\lambda (\tau,-{\mathbf k}) \, \hat b^{\dagger}_\lambda(-\mathbf k) \right] \, ,
\end{equation}
where we have employed the decomposition in terms of helicity (or chiral) states $\lambda = R, L$. For the purpose of what follows, we are decomposing tensor perturbations in terms of the helicity polarization basis defined through 
\begin{align} \label{eq:helicity_tensor_pol}
e_{ij}^{R, L} &= \frac{1}{\sqrt 2} \left[e_{ij}^+ \pm i \, e_{ij}^\times \right] \, ,\\
h_{R, L} &= \frac{1}{\sqrt 2} \left[ h_+ \mp i \, h_\times \right] \, ,
\end{align}
where $h_{+,\times}$ and $e_{ij}^{+, \times}$ are the usual linear polarizations of tensor perturbations. 

We remind that, if the tensor wave-vector is written in polar coordinates as 
\begin{equation}
\hat k = (\sin\varphi_1\cos\varphi_2,\sin\varphi_1\sin\varphi_2,\cos\varphi_1)\, ,
\end{equation}
we can define the linear polarization tensors in terms of two unit vectors perpendicular to $\hat k$ as
\begin{align} \label{eq:linear_tensor_pol}
e_{ij}^{+} &= (u_1)_i (u_1)_j - (u_2)_i (u_2)_j \, , \nonumber \\
e_{ij}^{\times} &= (u_1)_i (u_2)_j + (u_2)_i (u_1)_j \, ,
\end{align}
where 
\begin{align}
u_1 = \left(\sin \varphi_2 , - \cos \varphi_2, 0\right)\, ,  \qquad u_2 = \begin{cases} \left(\cos \varphi_1 \cos \varphi_2 , \cos \varphi_1 \sin \varphi_2, - \sin \varphi_1 \right) \qquad \mbox{if } \varphi_1 \leq \pi/2 \, ,\\
- \left(\cos \varphi_1 \cos \varphi_2 , \cos \varphi_1 \sin \varphi_2, - \sin \varphi_1 \right) \qquad \mbox{if } \varphi_1 > \pi/2 \, . \end{cases} 
\end{align}
The helicity polarization basis introduced is normalized such that it satisfies the following identities
\begin{align}
e_{ij}^{L}(\hat k) e_{L}^{ij}(\hat k)&=e_{ij}^{R}(\hat k) e_{R}^{ij}(\hat k) = 0 \, , \nonumber\\
e_{ij}^L(\hat k) e_R^{ij}(\hat k)&= 2 \, ,\nonumber\\
e_{ij}^{\lambda}(-\hat k)&= e_{ij}^{\lambda}(\hat k) \, ,\nonumber\\
e^{R *}_{ij}(\hat k)&= e^{L}_{ij}(\hat k),\nonumber\\
h^{R*}_{-\bk}&= h_{\bk}^{L}  \, ,\nonumber\\
k_l \, \bar \epsilon^{mlj} \, {e_j^{(\lambda) i}}(\hat k)&= \mp i \alpha_\lambda \, k \, e^{(\lambda) im}(\hat k) \, ,\label{eq:circ_identities}
\end{align}
where $\alpha_R=+1$ and $\alpha_L=-1$, and $\bar \epsilon^{mlj}$ denotes the Levi-Civita anti-symmetric symbol. Also, the $\mp$ in the last equation refers to the cases $\varphi_1 \leq$ and $> \pi/2$, respectively. In the following, we will assume $\varphi_1 \leq \pi/2$. The final predictions will not be sensitive to this choice.

At leading order in slow-roll and fixing the Bunch-Davies initial condition we find the following free solution for the tensor mode function 
\begin{equation} \label{eq:mode_function_tensor}
u^h_\lambda (\tau, k) =  \frac{H}{M_{pl}\sqrt{k^3}} \, (1+ i k \tau) \, e^{- i k \tau} \, ,
\end{equation}
which does not depend on the helicity $\lambda$. 

Contrary to the scalar sector, unless we introduce additional operators, we can not produce sizable tensor and scalar-tensor non-Gaussianities at tree level due to the slow-roll dynamics and the absence of a linear coupling of the form $\delta\sigma h$.

\subsection{Gravitational Chern-Simons operator coupled to a spectator field}

The gravitational Chern-Simons (gCS) operator can be expressed as the four-divergence of the gravitational Chern-Simons current $K^{\mu}$ as \cite{Jackiw:2003pm}
\begin{align} \label{eq:def_CS}
^*{R}^{(4)} R^{(4)} &=  2 \, \nabla_\mu K^\mu \, ,\\
K^\mu &= \epsilon^{\mu\nu\alpha\beta}\left(\Gamma^\sigma_{\nu\rho}\partial_\alpha \Gamma^\rho_{\beta\sigma} + \frac{2}{3}\Gamma^\sigma_{\nu\rho}\Gamma^\rho_{\alpha\lambda}\Gamma^\lambda_{\beta\sigma}\right) \, ,  \label{eq:def_K}
\end{align}
where $^*{R}^{(4)} R^{(4)}$ denotes the Riemann tensor contracted with its dual, $\epsilon^{\mu\nu\alpha\beta}$ is the covariant Levi-Civita tensor and $\Gamma^\sigma_{\nu\rho}$ are the components of the Christoffel symbols. 

Since the gCS operator can be expressed as a four-divergence of a current, it is well known that this term has non-vanishing effects only if it is coupled to a scalar field. Moreover, it can be shown that we can re-write this term by replacing the Riemann tensor with the Weyl tensor on the l.h.s. of the definition \eqref{eq:def_CS} \cite{Weinberg:2008hq}. As the (FRW) background metric is conformally flat, the Weyl tensor is vanishing on the background, and hence the gCS operator does not affect the background equations of motion. 

Here we consider the gCS term during inflation coupled to the spectator field $\sigma$ through a coupling $f(\sigma)$. The resulting new operator reads
\begin{align}\label{eq:sigma_CS_coup}
\Delta S_{CS} &= \int d^4 x \sqrt{-g} \, f(\sigma) ^*{R}^{(4)} R^{(4)} = 2 \int d^4 x \sqrt{-g} \, f(\sigma) \nabla_\mu K^\mu = 2 \int d^4 x  \,f(\sigma) \partial_\mu (\sqrt{-g} K^\mu)  \nonumber \\
&= - 2 \int d^4 x \sqrt{-g}  \, (f'(\sigma)   K^0 + \partial_i f(\sigma) \, K^i) \, .
\end{align}
As a reference, in the following we will take a minimal coupling of the kind 
\begin{align} \label{eq:CS_coupling}
f(\sigma) = \kappa_{CS} \,\sigma \, ,
\end{align}
where $\kappa_{CS}>0$ denotes the $\delta\sigma$-gCS coupling constant. The total action of our theory is
\begin{equation} \label{eq:action_model}
S = S_{\rm kin} + \Delta S_{CS} \, .
\end{equation}
The additional gCS contribution does not modify the background equations, but it does affect primordial perturbations. In the conventional setup the gCS term is coupled to the inflaton field. In this case it is well-known that the background dynamics of the inflaton leads to a quadratic correction in tensor perturbations. Due to this modification, tensor perturbations of one of the two helicities become ghost-like in nature at wave-numbers larger than the so-called Chern-Simons mass (see e.g. \cite{Dyda:2012rj}) 
\begin{align}
M^\phi_{CS} = \frac{M^2_{pl}}{8\kappa_{CS} |\dot \phi_0|} \,.
\end{align}
To avoid the formation of ghost fields at the characteristic energy scale of inflation we need to impose 
\begin{align} \label{eq:CS_condition}
\frac{H}{M_{CS}} \ll 1 \, , 
\end{align}
which severally constrain the product $\kappa_{CS}|\dot \phi_0|$, and therefore the observability prospects of the theory both in the tensor power spectrum and higher-order correlators. Instead, when the gCS term is coupled to the spectator field, the Chern-Simons mass read
\begin{align} \label{eq:MCS_sigma}
M^\sigma_{CS} = \frac{M^2_{pl}}{8\kappa_{CS} |\dot \sigma_0|} \, , 
\end{align}
and only the corrections to the tensor power spectrum are attenuated, as they contain the combination $\kappa_{CS}|\dot \sigma_0|$. Conversely, the cubic Lagrangian contains terms sensitive only to $\kappa_{CS}$, and therefore if $|\dot \sigma_0|$ is small enough, sizable non-Gaussianities could be sourced without spoiling the condition~\eqref{eq:CS_condition}. We see this once we perturb the coupling $f(\delta\sigma) = \kappa_{CS} \delta\sigma$ and note that $K^\mu$ starts at quadratic order in the perturbations (see Eqs. \eqref{eq:kappa0_current} and \eqref{eq:kappai_current}). Therefore, the strength of the leading cubic order Lagrangian is controlled by $\kappa_{CS}$ only.

As computed above, in our multi-field model
\begin{align} \label{eq:conditions_sigmadotCS}
\dot\sigma_0  = \frac{\kappa_\sigma \dot\phi_0 \ddot\phi_0}{m^2_\sigma} = \frac{2\kappa_\sigma \epsilon |\epsilon-\eta| M^2_{pl} H^3}{m^2_\sigma}\, . 
\end{align}
In Eq. \eqref{eq:conditions_sigmadotCS} we have expressed everything in terms of the leading slow-roll parameters. Therefore, by plugging Eq.~\eqref{eq:conditions_sigmadotCS} into \eqref{eq:MCS_sigma} and imposing the constraint in Eq.~\eqref{eq:CS_condition}, we find the following constraint on the product $\kappa_{CS} \kappa_\sigma$:
\begin{align} \label{eq:conditions_sigmadotCS2}
\kappa_{CS} \kappa_\sigma \ll \frac{m_\sigma^2}{H^2}\frac{10^4\epsilon^{-3/2} |\epsilon-\eta|^{-1}}{16 H M_{pl}} \, , 
\end{align}
where we have used 
\begin{align} \label{eq:HvsMpl}
H \simeq 10^{-4} \sqrt \epsilon \, M_{pl}\, . 
\end{align}
We will return later on in Sec. \ref{sec:pert_bound} on this bound. One may ask whether a more relaxed constraint than Eq. \eqref{eq:conditions_sigmadotCS2} can be obtained by considering an alternative $\sigma$-self potential. By applying the time derivative to Eq. \eqref{eq:centr_compensated_V} we find the more general equation
\begin{align} 
\dot\sigma_0  = \frac{\kappa_\sigma \dot\phi_0 \ddot\phi_0}{m^2_{\sigma, \rm eff}} = \frac{2\kappa_\sigma \epsilon |\epsilon-\eta| M^2_{pl} H^3}{m^2_{\sigma, \rm eff}} \, ,
\end{align}
where $m^2_{\sigma, \rm eff} = \partial^2_\sigma V(\sigma_0)$ is an effective mass which would also appear in Eq. \eqref{eq:quadratic_sigma}. Therefore, we would obtain the same constraint as in Eq. \eqref{eq:conditions_sigmadotCS2} with $m_{\sigma, \rm eff}$ replacing $m_{\sigma}$.

To conclude this section, we note that since the term in Eq. \eqref{eq:sigma_CS_coup} contains higher order derivatives of the metric with respect to the scalar curvature, second order time derivatives of the fields $h_{ij}$ and $\zeta$ may appear, as well as time derivatives of the auxiliary fields $N$ and $N_i$. However, as we are treating this additional term perturbatively, we can use the prescriptions of the effective field theory of inflation \cite{Weinberg:2008hq}, i.e. we can apply the time derivatives to the (first order) solutions for the fields $N$ and $N_i$ and express the second order time derivatives in terms of lower orders by the means of the zero-th order equations of motion for the perturbations
\begin{align}
\zeta'' + 2 a H \zeta' - \partial^2 \zeta &= 0 \, , \label{eq:zeroth_zeta}\\ 
h_{ij}'' + 2 a H h_{ij}' - \partial^2 h_{ij} &= 0 \, .
\end{align}
In the following we will perform the computation of the cubic interactions arising by the gCS operator coupled to $\sigma$. 

\section{Computation of the cubic Lagrangian}
\label{sec:cubic_comp}
Consider in the following the perturbed Friedmann-Lem\^{a}itre-Robertson-Walker (FLRW) metric in the ADM form 
\begin{align}
ds^2 &= g_{\mu\nu} dx^\mu dx^\nu = -(N^2 - N^i N_i \,) \, dt^2 + 2 N_i  \, dx^i dt + a^2 \, \tilde g_{ij} \, dx^i dx^j \, ,
\end{align}
where we factorized the scale factor in the spatial part of the metric as $\tilde g_{ij} =  g_{ij}/a^2$. By adopting the conformal time this metric reads
\begin{align}
ds^2 &= -(N^2 - N^i N_i) \, a^2 d\tau^2 + 2 N_i \, a \, dx^i d\tau + a^2 \, \tilde g_{ij} \, dx^i dx^j \, .
\end{align}
By re-defining 
\begin{align} \label{eq:redNi}
N_i = a \tilde N_i \, , 
\end{align}
we get 
\begin{align}
ds^2 &= -(N^2 - a^2\tilde N^i \tilde N_i) \, a^2 d\tau^2 + 2 \tilde N_i \, a^2 \, dx^i d\tau + a^2 \, \tilde g_{ij} \, dx^i dx^j \, .
\end{align}
Exploiting the fact that the gCS term is invariant under a conformal transformation of the metric \cite{Tian:2015vda}, we can compute it using the transformed metric $g'_{\mu \nu} = a^{-2} \, g_{\mu \nu}$, which reads
\begin{align} \label{eq:transf_metric}
ds^2 &= g'_{\mu\nu} dx^\mu dx^\nu = -(N^2 - a^2\tilde N^i \tilde N_i) \,  d\tau^2 + 2 \tilde N_i \, dx^i d\tau + \, \tilde g_{ij} \, dx^i dx^j \, .
\end{align} 
By employing this metric we can perform the Latin contractions with the metric $\tilde g_{ij}$ which is independent on the scale factor $a$. 
The inverse metric reads
\begin{align} \label{eq:Inverse_transf_metric}
ds^2 & = g'^{\mu\nu} dx_\mu dx_\nu = - N^{-2} \, d\tau^2 + 2 a^2 \tilde N^i N^{-2}  \, dx_i d \tau + \, \left(\tilde g^{ij}- a^2\frac{\tilde N^i \tilde N^j}{N^2}\right) \, dx_i dx_j \, .
\end{align}
As shown e.g. in \cite{Maldacena:2002vr}, when computing the cubic interaction Lagrangian we only need the value of the auxiliary fields $N$ and $\tilde N_i$ up to first order in terms of the perturbations $\zeta$, $\delta\sigma$ and $h_{ij}$. These can be read in Eq. \eqref{eq:constraints_quasi} once we remember the field re-definition \eqref{eq:redNi} and we switch to conformal time
\begin{align} 
N &= 1+\frac{\zeta'}{aH} \, , \nonumber \\
\tilde N_i &= \partial_i \psi \, , \qquad\qquad \psi = - \frac{\zeta}{a H} + \chi \, , \qquad\qquad \partial^2\chi = a \epsilon \left(\frac{\zeta'}{a} - \frac{1}{2}H \kappa_\sigma \,\delta \sigma\right) \, . 
\end{align}
For simplicity of notation from now on we will refer to $N$ to indicate its perturbed part only.

The Christoffel symbols $\Gamma^\mu_{\alpha\beta}$ read
\begin{equation}
\Gamma^\mu_{\alpha\beta} = \frac{g^{\mu\lambda}}{2}\left(\partial_\alpha g_{\lambda\beta} + \partial_\beta g_{\alpha\lambda }  - \partial_\lambda g_{\alpha\beta}\right) \, ,
\end{equation}
and their background value is 0 when computed with the metric \eqref{eq:transf_metric} since the conformal transformation performed erased the scale factor from the background metric. From Eq.~\eqref{eq:sigma_CS_coup} we read that the cubic action piece which does not contain dependence over $\dot\sigma_0$ comes from the computation of the 4-current $K^\mu$ at second order in the perturbations multiplying the perturbed coupling $f(\delta\sigma)$. From the definition \eqref{eq:def_K}, which we formally rewrite
\begin{align}
K \sim \Gamma \partial \Gamma + \Gamma \Gamma \Gamma \, , 
\end{align}
it follows that only the term $\propto \Gamma^{(1)} \partial \Gamma^{(1)}$ contributes, where the suffixes refer to the order of each term in perturbation theory. Therefore, we need to compute the Christoffel symbols only up to first order in the perturbations
\begin{align} \label{eq:Christ_first}
{\Gamma^\mu_{\alpha\beta}} = &\frac{\left[g^{\mu\lambda}\right]^{(0)}}{2} \left[\left(\partial_\alpha g_{\lambda\beta} + \partial_\beta g_{\alpha\lambda}  - \partial_\lambda g_{\alpha\beta}\right)\right]^{(1)}  + \mbox{(quadratic)} \, .
\end{align}
We employ the co-moving gauge \eqref{eq:comoving_gauge} and expand in perturbations the metric components in Eqs. \eqref{eq:transf_metric} and \eqref{eq:Inverse_transf_metric} up to the desired order:
\begin{align} \label{eq:metric}
 & g_{00} = -1 - 2 N + ...\, , \qquad g_{0i} = g_{i0} =  
\tilde N_i + ... \, , \qquad  g_{ij} = \delta_{ij} (1+2\zeta) + h_{ij} +  ... \, , \nonumber \\
 & g^{00} = -1 +...\, , \qquad g^{0i} = g^{i0} =  0 + ... \, , \qquad  g^{ij} = \delta^{ij} + ... \, .
\end{align}
Using these metric components we get the following expression for the Christoffel symbols
\begin{align}
\Gamma^0_{00} &= N' + \mbox{(quadratic)} \, ,\\
\Gamma^0_{0i} &= \partial_i N  + \mbox{(quadratic)} \, ,\\
\Gamma^{0}_{ij} &= -\frac{1}{2} \partial_i \tilde N_j - \frac{1}{2}\partial_j \tilde N_i + \frac{1}{2} h_{ij}' + \delta_{ij} \zeta' + \mbox{(quadratic)} \, ,  \\
\Gamma^i_{00} &= \partial_i N + \tilde N'_i + \mbox{(quadratic)} \, ,  \\
\Gamma^i_{0j} &=  \frac{1}{2} \partial_j \tilde N_i - \frac{1}{2} \partial_i \tilde N_j + \frac{1}{2} h'_{ij} + \delta_{ij} \zeta' + \mbox{(quadratic)} \, , \\
\Gamma^i_{jk} &= \frac{1}{2} \partial_k h_{ij} + \frac{1}{2} \partial_j h_{ik} - \frac{1}{2} \partial_i h_{jk} + \delta_{ik} (\partial_j \zeta) + \delta_{ij} (\partial_k \zeta) - \delta_{jk} (\partial_i \zeta) + \mbox{(quadratic)} \, . 
\end{align}
Plugging these in Eq. \eqref{eq:def_K} and doing some algebra we get the Chern-Simons 4-current at quadratic order
\begin{align} \label{eq:kappa0_current}
K^0 &= \epsilon^{0ijk}\left(\Gamma^\sigma_{i\rho}\partial_j\Gamma^\rho_{k\sigma}\right) = \epsilon^{0ijk}\left(\Gamma^0_{i0}\partial_j\Gamma^0_{k0} + \Gamma^0_{il}\partial_j\Gamma^l_{k0} + \Gamma^l_{i0}\partial_j\Gamma^0_{kl} + \Gamma^l_{im}\partial_j\Gamma^{m}_{kl}\right) \nonumber \\
&= \epsilon^{ijk}\Bigg{\{} (\partial_j\partial_l \zeta) (\partial_k h_{il}) - \frac{1}{2}\left (\partial_i \tilde N_l\right) \left(\partial_j h'_{lk}\right) + \frac{1}{2} h'_{il}\left(\partial_j h'_{lk}\right) + \frac{1}{2} \left(\partial_m h_{il}\right) \left(\partial_j\partial_l h_{km}\right) \nonumber \\
&\qquad\qquad\qquad\qquad\qquad\qquad\qquad\qquad\qquad\qquad\qquad - \frac{1}{2} \left(\partial_m h_{il}\right) \left(\partial_j\partial_m h_{kl} \right) \Bigg{\}} + \mbox{(cubic)} \, ,\\
K^i &= \epsilon^{ij0k}\left(\Gamma^\sigma_{j\rho}\partial_0\Gamma^\rho_{k\sigma}\right) + \epsilon^{i0jk}\left(\Gamma^\sigma_{0\rho}\partial_j\Gamma^\rho_{k\sigma}\right) + \epsilon^{ijk0}\left(\Gamma^\sigma_{j\rho}\partial_k\Gamma^\rho_{0\sigma}\right) \nonumber \\
&= \epsilon^{0ijk}\left(\Gamma^0_{j0}\partial_0\Gamma^0_{k0} + \Gamma^0_{jl}\partial_0\Gamma^l_{k0} + \Gamma^l_{j0}\partial_0\Gamma^0_{kl} + \Gamma^l_{jm}\partial_0\Gamma^{m}_{kl}\right)  \nonumber\\
&\qquad\qquad\qquad\qquad - 2 \epsilon^{0ijk} \left(\Gamma^0_{00}\partial_j\Gamma^0_{k0} + \Gamma^l_{00}\partial_j\Gamma^0_{kl} + \Gamma^0_{0l}\partial_j\Gamma^l_{k0} + \Gamma^l_{0m}\partial_j\Gamma^{m}_{kl}\right) \nonumber\\
&= \epsilon^{ijk}\Bigg{\{}  - \frac{1}{2} \left(\partial_l \tilde N_j\right) h''_{lk} - \frac{1}{2}  h'_{lj} \left(\partial_l \tilde N'_k\right) + (\partial_m \zeta) (\partial_j h'_{km}) + (\partial_k h_{jm}) (\partial_m \zeta')   \nonumber \\
& \qquad\qquad\qquad\qquad\qquad\qquad\quad + \frac{1}{2}  h'_{jl} h''_{lk} + \frac{1}{2} \left(\partial_m h_{jl}\right) \left(\partial_l h'_{km}\right) - \frac{1}{2} \left(\partial_m h_{jl}\right) \left(\partial_m h'_{lk}\right) \Bigg{\}} \nonumber \\
& \qquad\qquad\qquad + \epsilon^{ijk}\Bigg{\{}  - 2 \left(\partial_l N\right) \left(\partial_j h'_{lk}\right) - \tilde N'_l \left(\partial_j h'_{lk}\right) \Bigg{\}}  + \mbox{(cubic)} \, , \label{eq:kappai_current}
\end{align}
where Latin indexes are contracted with the Kronecker-delta. By reminding the relationship between the covariant Levi-Civita tensor and the Levi-Civita symbol $\bar \epsilon^{ijk}$:
\begin{align}
\bar \epsilon^{ijk} = \sqrt{-g} \, \epsilon^{ijk} \, , 
\end{align}
the final cubic Lagrangian can be obtained by contracting the $K^\mu$ current components with the derivatives of the Taylor expanded coupling function in Eq. \eqref{eq:CS_coupling} 
\begin{align} \label{eq:expand_S}
\Delta S_{CS} = - 2 \int d^4 x \, \Big[ \kappa_{CS} \, \delta \sigma' \, K^0 +  \kappa_{CS} \, (\partial_i \delta \sigma)  K^i \Big] \, , 
\end{align}
with $\bar \epsilon^{ijk}$ replacing the covariant Levi-Civita tensor in Eqs. \eqref{eq:kappa0_current} and \eqref{eq:kappai_current}. 

By collecting all together in Eq. \eqref{eq:expand_S} and doing appropriate integration by parts we get the following cubic interaction Lagrangian 
\begin{align} \label{eq:action_expanded}
\left[\Delta \mathcal L_{CS}\right]^{(3)} = & - 2 \kappa_{CS} \, \bar\epsilon^{ijk} \Big[ - 2 \delta\sigma (\partial_i\partial_l \zeta) (\partial_j h'_{lk}) + 2 \delta\sigma (\partial_i\partial_l N) (\partial_j h'_{lk}) + 2 \delta\sigma (\partial_i \tilde N'_l) (\partial_j h'_{lk}) \nonumber \\
&\qquad\qquad\qquad\qquad  - \delta\sigma h''_{il} (\partial_j h'_{lk}) - \delta\sigma (\partial_m h'_{il}) (\partial_j\partial_l h_{km}) + \delta\sigma (\partial_m h'_{il}) (\partial_j\partial_m h_{kl}) \Big] \, . 
\end{align}
As anticipated, we note that second order time derivatives appear, while at least at cubic order the fields $N$ and $\tilde N_i$ remain Lagrange multipliers\footnote{Even if the third term in Eq. \eqref{eq:action_expanded} contains a time derivative of $\tilde N_i$, we can always remove it by performing an integration by parts.}. Note that in principle terms who depend on the auxiliary fields will provide additional contributions to the equations of motion of $N$ and $\tilde N_i$. However, these corrections arise at second order in the perturbations and therefore we can neglect them in the computation of the cubic Lagrangian. By substituting the equations for the constraints we get
\begin{align}
\left[\Delta \mathcal L_{CS}\right]^{(3)} &= - 2 \kappa_{CS} \, \bar\epsilon^{ijk} \nonumber\\
&\times \Big[ 2 \epsilon\, \delta\sigma (\partial_i\partial_l \partial^{-2} \zeta'') (\partial_j h'_{lk}) - 2 \epsilon\, \delta\sigma (\partial_i\partial_l \zeta) (\partial_j h'_{lk}) \nonumber \\
&\qquad\qquad- \epsilon \kappa_{\sigma} a H \, \delta\sigma (\partial_i\partial_l \partial^{-2} \delta\sigma') (\partial_j h'_{lk}) - \epsilon \kappa_{\sigma} a^2 H^2  \, \delta\sigma (\partial_i\partial_l \partial^{-2} \delta\sigma) (\partial_j h'_{lk})  \nonumber \\
& \qquad\qquad - \delta\sigma h''_{il} (\partial_j h'_{lk}) - \delta\sigma (\partial_m h'_{il}) (\partial_j\partial_l h_{km}) + \delta\sigma (\partial_m h'_{il}) (\partial_j\partial_m h_{kl}) \Big] \, . 
\end{align}
Now, we use the zero-th order equations of motion for $\zeta$ and $h_{ij}$ to eliminate second order time derivatives
\begin{align}
\left[\Delta \mathcal L_{CS}\right]^{(3)} &= - 2 \kappa_{CS}  \, \bar\epsilon^{ijk} \nonumber \\
& \times \Big[ -4\epsilon a H \, \delta\sigma (\partial_i\partial_l \partial^{-2} \zeta') (\partial_j h'_{lk}) - \epsilon \kappa_{\sigma} a H \, \delta\sigma (\partial_i\partial_l \partial^{-2} \delta\sigma') (\partial_j h'_{lk}) \nonumber \\ 
& \qquad\qquad - \epsilon \kappa_{\sigma} a^2 H^2\, \delta\sigma (\partial_i\partial_l \partial^{-2} \delta\sigma) (\partial_j h'_{lk})  + 2 a H \,\delta\sigma h'_{il} (\partial_j h'_{lk}) - \delta\sigma (\partial^2 h_{il}) (\partial_j h'_{lk}) \nonumber \\
&\qquad\qquad\qquad\qquad\qquad\qquad\quad- \delta\sigma (\partial_m h'_{il}) (\partial_j\partial_l h_{km}) + \delta\sigma (\partial_m h'_{il}) (\partial_j\partial_m h_{kl}) \Big] \, . 
\end{align}
From this equation we can distinguish the following cubic interactions 
\begin{align}
\left[\Delta \mathcal L_{CS}\right]^{(3)}_{\sigma hh} = 2 \kappa_{CS} \, \bar\epsilon^{ijk} \Big[ & -2 a H \,\delta\sigma h'_{il} (\partial_j h'_{lk}) + \delta\sigma (\partial^2 h_{il}) (\partial_j h'_{lk}) \nonumber\\
& \qquad+\delta\sigma (\partial_l h'_{im}) (\partial_j\partial_m h_{lk}) - \delta\sigma (\partial_m h'_{il}) (\partial_j\partial_m h_{lk}) \Big] \, , 
\end{align}
\begin{align} 
&\left[\Delta \mathcal L_{CS}\right]^{(3)}_{\sigma\zeta h} = 8 a H \epsilon \kappa_{CS} \, \bar\epsilon^{ijk} \, \delta\sigma (\partial_i\partial_l \partial^{-2} \zeta') (\partial_j h'_{lk})  \, , 
\end{align}
and 
\begin{align} 
&\left[\Delta \mathcal L_{CS}\right]^{(3)}_{\sigma\sigma h} =    2\epsilon \kappa_{CS} \kappa_{\sigma}\, \bar\epsilon^{ijk} \Big[ a H \, \delta\sigma (\partial_i\partial_l \partial^{-2} \delta\sigma') (\partial_j h'_{lk}) + a^2 H^2 \, \delta\sigma (\partial_i\partial_l \partial^{-2} \delta\sigma) (\partial_j h'_{lk}) \Big] \, . 
\end{align}
These interactions in combination with the exchange vertex \eqref{eq:intst} can provide additional (parity-violating) tree-level contributions to $\langle \zeta \zeta h \rangle$ and $\langle \zeta h h \rangle$.  

For the purpose of the computations we will perform in the next section, it is convenient to convert the interactions in Fourier space, and use the last equation of Eqs. \eqref{eq:circ_identities} and $a \simeq - 1/H\tau$. We get
\begin{align} \label{eq:hhsigma}
\left[\Delta \mathcal L_{CS}\right]^{(3)}_{\sigma h_{\lambda} h_{\tilde\lambda}} =  - 2   \kappa_{CS} \, \alpha_{\tilde\lambda}  \Big[ & -\frac{2 p_3}{\tau} \, \delta\sigma(\mathbf{p}_1) \, h'_{\lambda}(\mathbf{p}_2) h'_{\tilde\lambda}(\mathbf{p}_3) \left(e^{\lambda}_{ij}(\hat{p}_2) \cdot e^{\tilde\lambda}_{ij}(\hat{p}_3)\right) \nonumber \\
&\quad + p_3 \,p^2_2 \, \delta\sigma(\mathbf{p}_1) \, h_{\lambda}(\mathbf{p}_2) h'_{\tilde\lambda}(\mathbf{p}_3) \left(e^{\lambda}_{ij}(\hat{p}_2) \cdot e^{\tilde\lambda}_{ij}(\hat{p}_3)\right) \nonumber\\
&\quad+ p_3 \, \, \delta\sigma(\mathbf{p}_1) \, h'_{\lambda}(\mathbf{p}_2) h_{\tilde\lambda}(\mathbf{p}_3) \left(p^l_2 \,p^m_3 \cdot e^{\lambda}_{im}(\hat{p}_2) e^{\tilde\lambda}_{il}(\hat{p}_3) \right) \nonumber \\
&\quad- p_3 \,(\mathbf{p}_2 \cdot \mathbf{p}_3) \, \delta\sigma(\mathbf{p}_1) \, h'_{\lambda}(\mathbf{p}_2) h_{\tilde\lambda}(\mathbf{p}_3) \left(e^{\lambda}_{ij}(\hat{p}_2) \cdot e^{\tilde\lambda}_{ij}(\hat{p}_3)\right) \Big] \, , 
\end{align}
\begin{align} \label{eq:hzetasigma}
\left[\Delta \mathcal L_{CS}\right]^{(3)}_{\sigma \zeta h_{\lambda}} = - 8 \epsilon \,  \kappa_{CS} \, \alpha_{\lambda} \Big[ \, & \frac{1}{\tau} \, \frac{p_3}{p_2^2} \delta\sigma(\mathbf{p}_1) \, \zeta'(\mathbf{p}_2) \, h'_{\lambda}(\mathbf{p}_3) \, \left(p^i_2 p^l_2 \cdot e^{\lambda}_{il}(\hat{p}_3)\right) \Big] \, , 
\end{align}
and  
\begin{align} \label{eq:hsigmasigma}
&\left[\Delta \mathcal L_{CS}\right]^{(3)}_{\sigma\sigma h_{\lambda}} = - 2 \epsilon \kappa_{CS} \kappa_{\sigma} \, \alpha_{\lambda}  \Big[ \frac{1}{\tau} \, \frac{p_3}{p_2^2} \delta\sigma(\mathbf{p}_1) \, \delta\sigma'(\mathbf{p}_2) \, h'_{\lambda}(\mathbf{p}_3) \, \left(p^i_2 p^l_2 \cdot e^{\lambda}_{il}(\hat{p}_3)\right) \nonumber \\
&\qquad \qquad \qquad \qquad \qquad \qquad \qquad \qquad \qquad - \frac{1}{\tau^2}  \,\frac{p_3}{p_2^2} \delta\sigma(\mathbf{p}_1) \, \delta\sigma(\mathbf{p}_2) \, h'_{\lambda}(\mathbf{p}_3) \, \left(p^i_2 p^l_2 \cdot e^{\lambda}_{il}(\hat{p}_3)\right) \Big] \, . 
\end{align}
Here the tensor perturbations have been decomposed in terms of their polarization tensor $e^{\lambda}_{ij}$ and their mode-functions $h_{\lambda}$, and the sum over the polarization indexes is understood. From these equations, we see the appearance of the helicity-dependent factor $\alpha_{\lambda}$, which explicitly signals the breaking of parity symmetry. As a result, scalar perturbations mix in a different way with tensor perturbations depending on the helicity of the tensor perturbations in the interaction.  

\section{Parity violation in scalar-tensor bispectra}
\label{sec:Feynman}

\subsection{Schwinger-Keldysh formalism}

In this section, we compute the contributions to the scalar-tensor bispectra arising from the interaction vertices introduced previously. To perform this computation, we adopt the Feynman diagrammatic rules of the Schwinger-Keldysh formalism, as described in~\cite{Chen:2017ryl} and originally proposed by Weinberg~\cite{Weinberg:2005vy}. We begin with a brief summary of the relevant aspects of this formalism.

The central principle of this method is to evaluate cosmological correlators analogously to scattering amplitudes. The primary components involved in the calculation are the interaction vertices—specifically Eqs.~\eqref{eq:intst}, \eqref{eq:hhsigma}, \eqref{eq:hzetasigma}, and \eqref{eq:hsigmasigma}—and the corresponding mode functions—namely Eqs.~\eqref{eq:uzeta}, \eqref{eq:vsigma_m}, and \eqref{eq:mode_function_tensor}.

Suppose we wish to compute the three-point correlation function of the fields $A_1(K_1^c)$, $A_2(K_2^c)$, and $A_3(K_3^c)$ at the time $\tau_0$ (referred to as the \emph{boundary} time), where the triplet $(K_1^c, K_2^c, K_3^c)$ denotes an independent combination of momenta $k_1$, $k_2$, and $k_3$. These fields correspond to the external lines in the associated Feynman diagrams. As per standard procedure, we enumerate all possible diagrams constructed from the available interaction vertices that connect the external lines. For each internal line, we assign a momentum fixed by momentum conservation at the vertices. We introduce additional momenta for each momentum not fixed by momentum conservation. Internal vertices are associated with times $\tau_i < \tau_0$ (denoted as \emph{bulk} times) and can carry either a plus $+$ or minus $-$ sign. The full result is obtained by summing over all combinations of independent external momenta (together with their polarization index if present) and vertex sign assignments. For each fixed configuration, the contribution of the corresponding Feynman diagram is determined by the following rules:

\begin{itemize}
    \item \textbf{Vertex Factors:}  
    Each $\pm$ vertex contributes a factor of $\mp i$, respectively. Additionally, for each vertex, one must include the opposite of the coefficient of the corresponding interaction term in the Lagrangian (expressed in Fourier space), evaluated at the time of the vertex and using the momenta of the interacting fields.

    \item \textbf{Propagators:}  
     Each line coming out from a vertex and the external lines are associated with field operators. Propagators arise from contractions between these lines. Each field operator can be decomposed as
    \begin{align*}
    A_i \sim u_i a_i + u_i^* a_i^\dagger \, ,
    \end{align*}
    where $u_i$ are the mode functions, and $a_i^\dagger$, $a_i$ are the creation and annihilation operators, respectively. Each field operator is evaluated at the momentum carried by the corresponding line. The following rules apply for contractions:
    \begin{itemize}
        \item[$\triangle$] For a contraction between a $-$ vertex at time $\tau_i$ and an external line at $\tau_0$, the contribution is
        \begin{align*}
        \langle 0| A_i(\tau_i) A_j(\tau_0) |0\rangle = u_i(\tau_i) u_j^*(\tau_0) \, .
        \end{align*}
        \item[$\triangle$] For a contraction between a $+$ vertex at \( \tau_i \) and an external line at \( \tau_0 \), the contribution is
        \begin{align*}
        \langle 0| A_j(\tau_0) A_i(\tau_i) |0\rangle = u_i^*(\tau_i) u_j(\tau_0) \, .
        \end{align*}
        \item[$\triangle$] For a contraction between a $+$ vertex at $ \tau_i$ and a $-$ vertex at $\tau_j$, the contribution is
        \begin{align*}
        \langle 0| A_i(\tau_i) A_j(\tau_j) |0\rangle = u_i^*(\tau_i) u_j(\tau_j) \, .
        \end{align*}
        \item[$\triangle$] For a contraction between two $+$ vertices, the contribution is the time-ordered correlator:
        \begin{align*}
        \langle 0| T[A_i(\tau_i) A_j(\tau_j)] |0\rangle = u_i^*(\tau_i) u_j(\tau_j) \Theta(\tau_j - \tau_i) + u_i(\tau_i) u_j^*(\tau_j) \Theta(\tau_i - \tau_j) \, ,
        \end{align*}
        where $\Theta(x)$ is the Heaviside function. 
        \item[$\triangle$] For a contraction between two $-$ vertices, the contribution is the anti-time-ordered correlator:
        \begin{align*}
        \langle 0| \bar{T}[A_i(\tau_i) A_j(\tau_j)] |0\rangle = u_i^*(\tau_i) u_j(\tau_j) \Theta(\tau_i - \tau_j) + u_i(\tau_i) u_j^*(\tau_j) \Theta(\tau_j - \tau_i) \, .
        \end{align*}
    \end{itemize}

    If a vertex involves a time derivative of a field, i.e., $A'_i(\tau_i)$, the above expressions remain valid, with the corresponding time derivative applied to the mode function.

    \item \textbf{Integration and final factors:}  
    The final result includes integration over all internal times $\tau_i$ (ranging from $-\infty$ to $\tau_0$) and over all unconstrained momenta. A delta function factor $(2\pi)^3 \delta^{(3)}\left(\sum_i \mathbf{k}_i\right)$ must also be included to enforce momentum conservation at each vertex.
\end{itemize}

\subsection{Computation of the scalar-scalar-tensor bispectrum}

\begin{figure}
        \centering
        \includegraphics[width=.49\textwidth]{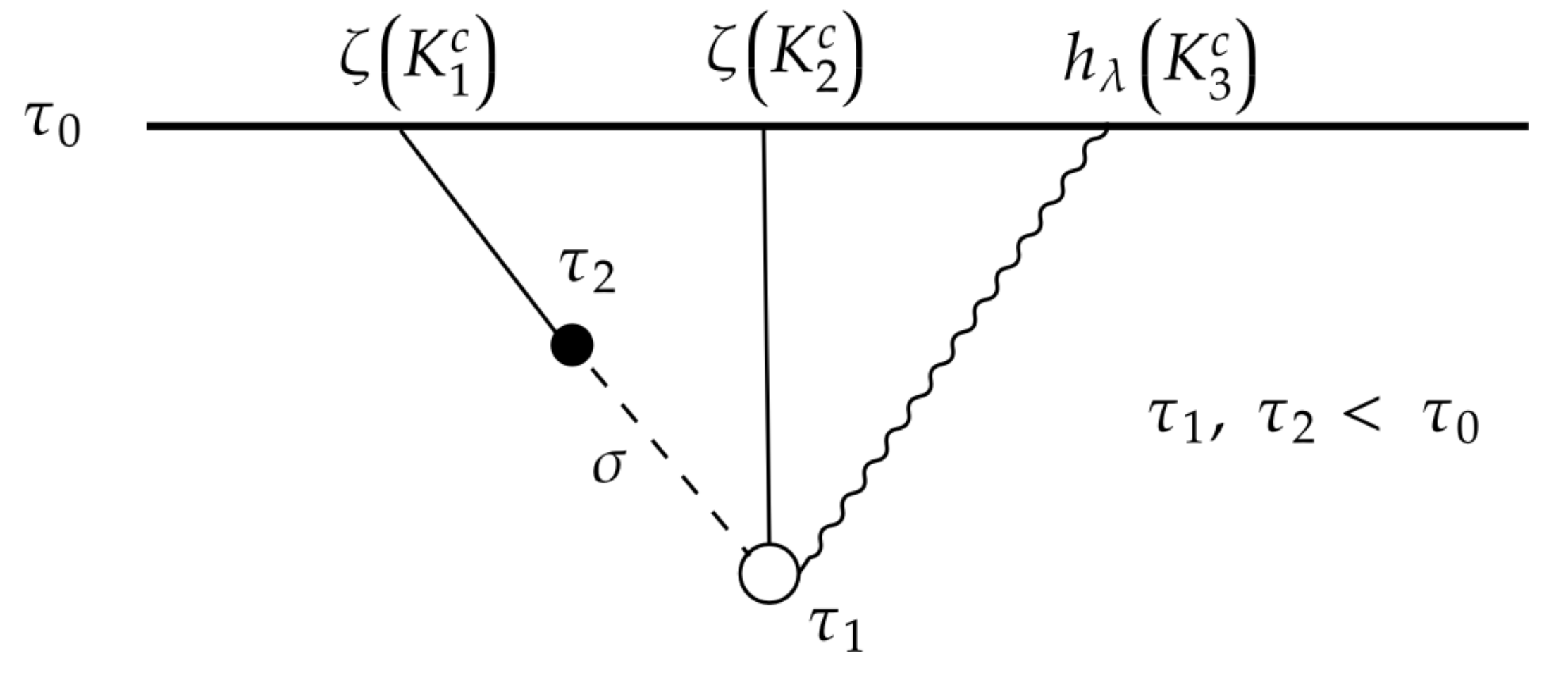} 
        \includegraphics[width=.49\textwidth]{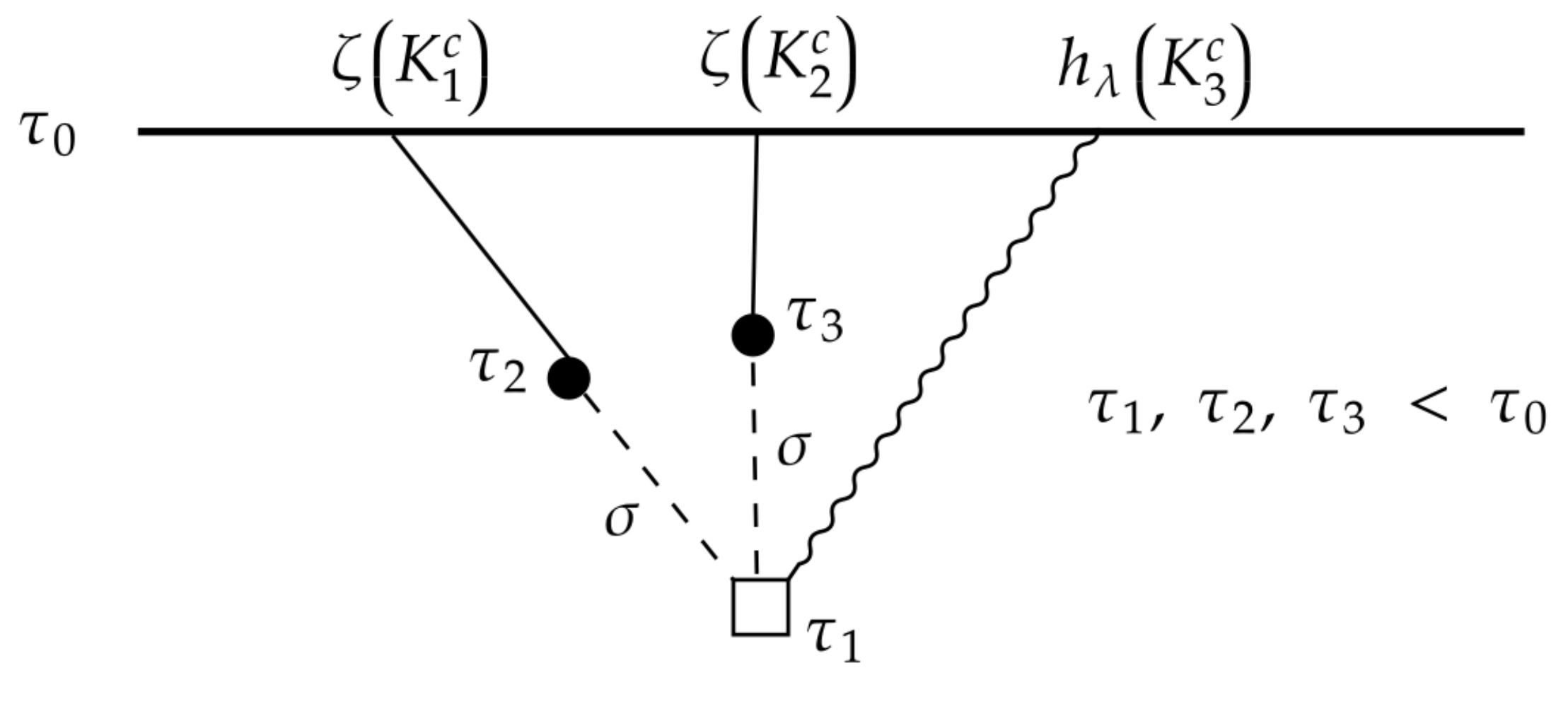}  
    \caption{Feynman diagrams for the scalar-scalar-tensor (parity-odd) bispectrum. The black vertex represents the $\sigma\zeta$ mediation in Eq. \eqref{eq:intst}. Left diagram: the white-round vertex represent the $\sigma \zeta h_\lambda$ mediation in Eq. \eqref{eq:hzetasigma}. Right diagram: the white-square vertex represents the $\sigma\sigma h_\lambda$ mediations in Eq. \eqref{eq:hsigmasigma}.} \label{fig:sst_diagrams_sigma_med}
\end{figure}

The Feynman diagrams for the computation of the scalar-scalar-tensor bispectrum are given in Fig. \ref{fig:sst_diagrams_sigma_med}. Given the structure of these diagrams and how the corresponding interaction vertices depend on the parameters of the theory, we can determine the scaling behavior of the final results as follows:
\begin{align}
\mbox{Left diagram} &\sim \frac{H^4}{M^4_{pl}} (H \kappa_{\sigma}) \, (H \kappa_{CS}) \, S_1(k_i) \, , \\
\mbox{Right diagram} &\sim \frac{H^4}{M^4_{pl}} \,\epsilon \, (H \kappa_{\sigma}) (M_{pl} \kappa_{\sigma})^2 \, (H \kappa_{CS})  \, S_2(k_i) \, , 
\end{align}
where $S_j(k_i)$ are functions of momenta that can be determined only by the full computation. From these scalings we can compute the ratio between the contribution from the right and left diagrams as 
\begin{align} \label{eq:RoverL}
\frac{\mbox{Right diagram}}{\mbox{Left diagram}} &\sim \epsilon \, (M_{pl} \kappa_{\sigma})^2  \, .
\end{align}
Note that this quantity must be small due to the perturbativity bound in Eq. \eqref{eq:radiative_control_mass_precise}. It follows that contributions from the left diagram, in principle, dominate over the ones from the right diagram. Therefore, in the following subsections we can limit ourselves to compute the effects coming from the $\sigma \zeta h$ interaction vertex derived in Eq. \eqref{eq:hzetasigma}. 

The Feynman diagram for the computation is given in Fig. \eqref{fig:sst_diagrams_sigma_med}, left panel. 

\begin{enumerate}
\item We label the external legs with the momenta $\bk_i$ and we consider all the possible independent combinations. Here in total we have two independent combinations obtaining exchanging the two scalar fields $\zeta$. In the following we write the two independent sets:
\begin{align}
\mathbf{K^{c_1}} = \{ \bk_1,\bk_2,\bk_3\} \, , \qquad\qquad \mathbf{K^{c_2}} = \{ \bk_2,\bk_1,\bk_3 \} \, . 
\end{align}
\item For each channel we label each vertex with either a $+$ and $-$. Therefore, in total there will be the following four combinations $\tau_1\tau_2$: $++$, $-+$, $+-$, $--$. We integrate over the time $\tau_1$ in the white vertex and over the time $\tau_2$ in the black vertex. Each time-integration goes from $-\infty$ to $\tau_0$, where here $\tau_0$ denotes the time at the end of inflation. For each $\pm$ vertex we multiply by $\mp i$, respectively. For each vertex we add the opposite of the corresponding coupling coefficient:
\begin{itemize}
\item black vertex (from Eq. \eqref{eq:intst}): $-2\frac{M^2_{pl}}{H^2} \frac{\epsilon \kappa_{\sigma}}{\tau^3}\, .$
\item white vertex (from Eq. \eqref{eq:hzetasigma}): $\frac{8}{\tau} \, \epsilon \kappa_{CS} \, \alpha_\lambda \frac{K^c_3}{(K_2^c)^2} \, \left(e^{\lambda}_{ij}(\hat K^c_3) \, (K_2^c)^i \, (K_2^c)^j \right) \, .$
\end{itemize}
\item For each scalar or tensor external leg connected to a $\pm$ vertex we have the following \textit{bulk-to-boundary} propagator
\begin{equation}
\mathcal{G'}_{\pm}(\tau, k) = \frac{H^2}{4 \epsilon k^3} \, k^2 \tau \,(1 \pm i k \tau_0) \,  e^{\pm i k \tau} \, e^{\mp i k \tau_0} \, .  
\end{equation} 
\item Then, depending on the vertex combination we have one of the following \textit{bulk} propagators for the spectator field $\delta\sigma$: 
\begin{align}
&++: \quad \Big[v_\sigma (\tau_1, K_1^c) \, v^{*}_\sigma (\tau_2, K_1^c) \, \Theta(\tau_1 - \tau_2) + v^*_\sigma (\tau_1, K_1^c) \, v_\sigma (\tau_2, K_1^c) \, \Theta(\tau_2 - \tau_1)\Big] \, , \\
&-+: \quad v_\sigma (\tau_1, K_1^c) \, v^{*}_\sigma (\tau_2, K_1^c) \, , \\
&+-: \quad v^*_\sigma (\tau_1, K_1^c) \, v_\sigma (\tau_2, K_1^c) \, , \\
&--: \quad  \Big[v^{*}_\sigma (\tau_1, K_1^c) \, v_\sigma (\tau_2, K_1^c) \, \Theta(\tau_1 - \tau_2) + v_\sigma (\tau_1, K_1^c) \, v^*_\sigma (\tau_2, K_1^c) \, \Theta(\tau_2 - \tau_1)\Big]  \, . 
\end{align}
Here the mode-function $v_\sigma$ corresponds to the one of the spectator field, Eq. \eqref{eq:vsigma_m}. 
\item In the end we need to sum over all the vertex combinations and channels (or momenta exchange) to obtain the final result. 
\end{enumerate}
By accounting for these rules, we get the final result\footnote{Here and afterwards we will omit the momentum conservation pre-factor $(2 \pi)^3 \delta^{(3)}(\bk_1+\bk_2+\bk_3)$.}
\begin{equation} \label{eq:sst_bisp}
\langle \zeta(\bk_1) \zeta(\bk_2) h_{\lambda}(\bk_3) \rangle =  - \alpha_\lambda \frac{H^4 \kappa_{CS} \kappa_{\sigma}}{M^4_{pl}} \sum_{c=c_1, c_2} \, \frac{1}{(K_1^c) (K_2^c)^3} \, \left(e^{\lambda}_{ij}(\hat K^c_3) \, (K_2^c)^i \, (K_2^c)^j \right) \, {\rm Tvar}^\nu_{1}(K^c_1,K^c_2,K_3^c) \, .
\end{equation}
In this result we factorize a momenta-dependent part and we need to compute a time-ordered variance
\begin{align} \label{eq:Tvar_GR1}
{\rm Tvar}^\nu_{1}(K^c_1,K^c_2,K_3^c) = - 2 \, {\rm Re}\left[I^{(1)}_{1}(K^c_1,K^c_2,K_3^c) + I^{(2)}_{1}(K^c_1,K^c_2,K_3^c) \right] \, ,
\end{align}
where 
\begin{align}\label{I1_comp}
I^{(1)}_{1}(K^c_1,K^c_2,K_3^c) =& \left(\prod_i (1 + i K^c_i \tau_0) \, e^{- i K^c_i \tau_0} \right) \nonumber \\
&\times \Big[ \int_{-\infty}^{\tau_0}  d \tau_1 \, \tau_1 \, v_\sigma (\tau_1, K_1^c) \, e^{i (K_2^c+K_3^c) \tau_1} \int_{-\infty}^{\tau_1}  \,  \frac{d \tau_2}{\tau^2_2} \, v^{*}_\sigma (\tau_2, K_1^c) \, e^{i K_1^c \tau_2}    \nonumber \\
& \qquad\qquad + \, \int_{-\infty}^{\tau_0}  d \tau_1 \, \tau_1  \, v^*_\sigma (\tau_1, K_1^c) \, e^{i (K_2^c + K_3^c) \tau_1} \int_{\tau_1}^{\tau_0} \frac{d \tau_2}{\tau^2_2} \,v_\sigma (\tau_2, K_1^c) \, e^{i K_1^c \tau_2} \Big] \, , 
\end{align}  
\begin{align} \label{I2_comp}
I^{(2)}_{1}(K^c_1,K^c_2,K_3^c) =& -(1 + i K^c_1 \tau_0) \, (1 - i K^c_2 \tau_0) \, (1 - i K^c_3 \tau_0) \, e^{- i (K^c_1-K^c_2-K^c_3) \tau_0} \nonumber \\
& \times \Big[ \int_{-\infty}^{\tau_0}  d \tau_1 \, \tau_1 \, v_\sigma (\tau_1, K_1^c)  \, e^{-i (K_2^c+K_3^c) \tau_1}  \int_{-\infty}^{\tau_0}  \frac{d \tau_2}{\tau^2_2} \, v^{*}_\sigma (\tau_2, K_1^c) \, e^{i K_1^c \tau_2} \Big]  \, . 
\end{align}
Assuming that the spectator field decays completely before the end of inflation, these time integrations can be computed analytically for a generic value of $0\leq\nu<3/2$ by taking the limit $\tau_0 \rightarrow 0$ (see Appendix \ref{app:time_var_comp} for the explicit computation). The final result reads
\begin{align} \label{eq:Tvar_GR1_computed}
{\rm Tvar}^\nu_{1}(K^c_1,K^c_2,K_3^c) =& -2 H^2 \pi \, \frac{1}{\left(2 K_1^c\right)^{4}} \, \frac{\Gamma(7/2-\nu) \Gamma(7/2+\nu)}{\Gamma(4)}  \nonumber \\
&\qquad\qquad\qquad\qquad\qquad \times {\rm H2F1}\left(7/2-\nu, \,7/2+\nu, \, 4, \,\frac{1}{2} - \frac{K_2^c+K_3^c}{2 K_1^c}\right)   \, ,
\end{align}
where $\Gamma(z)$ is the Gamma function and ${\rm H2F1}\left(a, \,b, \, c, \,z\right)$ denotes the Gauss hypergeometric function. 

When $\sigma$ is massless, $\nu = 3/2$, Eq. \eqref{eq:Tvar_GR1_computed} is no longer valid as the spectator field lasts until the end of inflation independently on the duration of inflation, and the correct result is found by substituting the mode-function of a massless field into Eqs. \eqref{I1_comp} and \eqref{I2_comp} and then doing the time-integrations analytically accounting for the finite duration of inflation. By doing so, we find 
\begin{align} \label{eq:Tvar1}
{\rm Tvar}^{\nu=3/2}_{1}(K^c_1,K^c_2,K_3^c) = 0 \, .
\end{align}

\subsection{Computation of the scalar-tensor-tensor bispectrum}

\begin{figure}
        \centering
        \includegraphics[width=.49\textwidth]{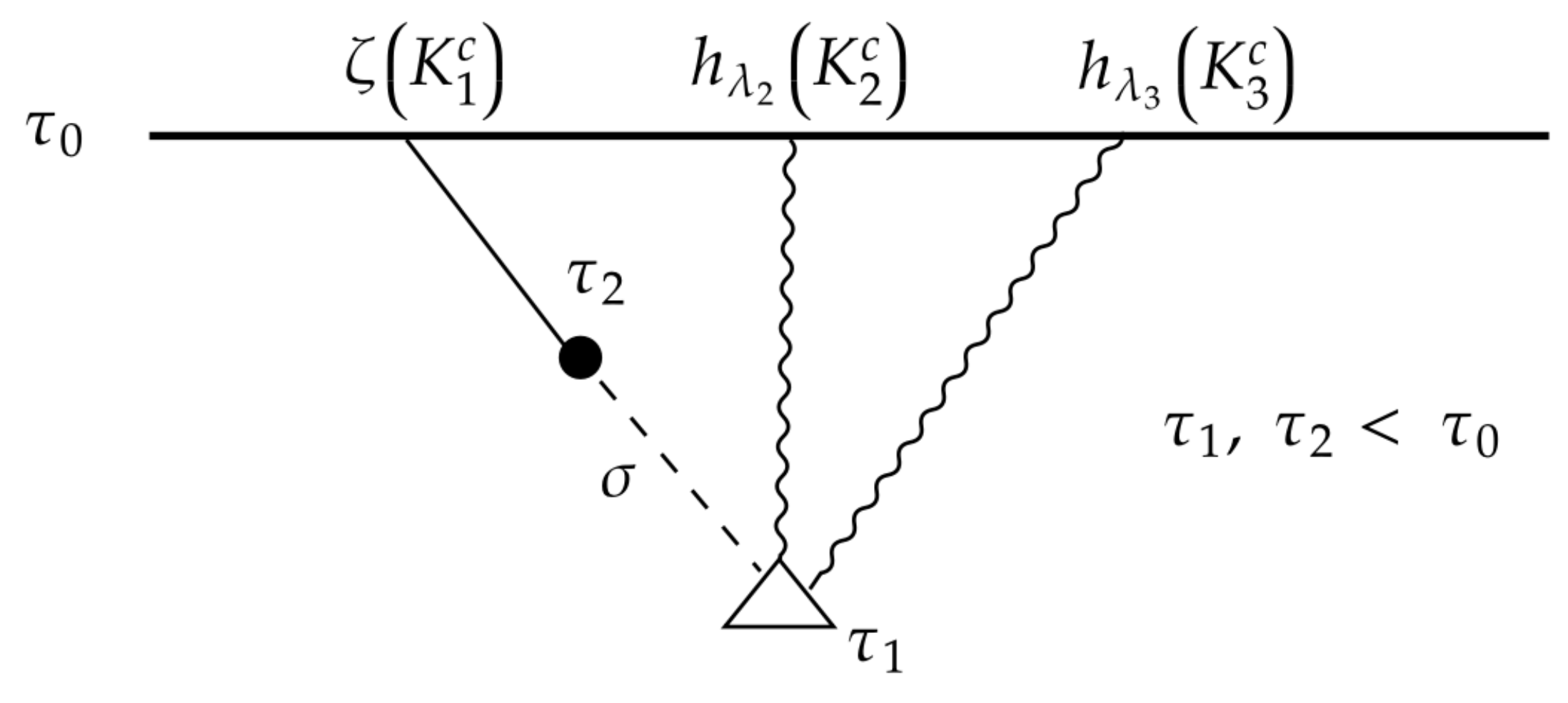}  
    \caption{Feynman diagram for the scalar-tensor-tensor (parity-odd) bispectrum. The black vertex represents the $\sigma\zeta$ mediation in Eq. \eqref{eq:intst}. The white-triangle vertex represents the $\sigma h_{\lambda_2} h_{\lambda_3}$ mediations in Eq. \eqref{eq:hhsigma}.} \label{fig:stt_diagrams_sigma_med}
\end{figure}

The Feynman diagram for the computation of the scalar-tensor-tensor bispectrum is illustrated in Fig. \ref{fig:stt_diagrams_sigma_med}. The corresponding contribution coming from the interactions in Eq. \eqref{eq:hhsigma} provides time-integrations that resemble the ones of the previous subsection, therefore here we do not provide all the intermediate steps.

In addition to the time-ordered variance in Eq. \eqref{eq:Tvar_GR1}, here we need to compute the following time-ordered variance 
\begin{align} \label{eq:Tvar_GR2}
{\rm Tvar}^\nu_{2}(K^c_1,K^c_2,K_3^c) = -2 \, {\rm Re}\left[I^{(1)}_{2}(K^c_1,K^c_2,K_3^c) + I^{(2)}_{2}(K^c_1,K^c_2,K_3^c) \right] \, ,
\end{align}
where
\begin{align}
I^{(1)}_{2}(K^c_1,K^c_2,K_3^c) =& \left(\prod_i (1 + i K^c_i \tau_0) \, e^{- i K^c_i \tau_0} \right) \nonumber \\
& \times \Big[\int_{-\infty}^{\tau_0}  d \tau_1 \, \tau_1 \, (1 - i K_2^c \tau_1) \,  v_\sigma (\tau_1, K_1^c) \, e^{i (K_2^c+K_3^c) \tau_1} \int_{-\infty}^{\tau_1}  \,  \frac{d \tau_2}{\tau^2_2} \, v^{*}_\sigma (\tau_2, K_1^c) \, e^{i K_1^c \tau_2}   \nonumber \\
& \qquad +  \, \int_{-\infty}^{\tau_0}  d \tau_1 \, \tau_1 \, (1 - i K_2^c \tau_1) \, v^*_\sigma (\tau_1, K_1^c) \, e^{i (K_2^c + K_3^c) \tau_1} \int_{\tau_1}^{\tau_0} \frac{d \tau_2}{\tau^2_2} \,v_\sigma (\tau_2, K_1^c) \, e^{i K_1^c \tau_2}\Big]  \, , 
\end{align} 
\begin{align}
I^{(2)}_{2}(K^c_1,K^c_2,K_3^c) =& -(1 + i K^c_1 \tau_0) \, (1 - i K^c_2 \tau_0) \, (1 - i K^c_3 \tau_0) \, e^{- i (K^c_1-K^c_2-K^c_3) \tau_0}  \nonumber \\
& \times \Big[\int_{-\infty}^{\tau_0}  d \tau_1 \, \tau_1 \, (1 + i K_2^c \tau_1) \, v_\sigma (\tau_1, K_1^c)  \, e^{-i (K_2^c+K_3^c) \tau_1} \,  \int_{-\infty}^{\tau_0}  \frac{d \tau_2}{\tau^2_2} \, v^{*}_\sigma (\tau_2, K_1^c) \, e^{i K_1^c \tau_2} \Big]   \, . 
\end{align}
As above, assuming that the spectator field completely decays before the end of inflation we can perform these integrations analytically for a generic value of $0\leq\nu<3/2$. The explicit computation resembles the one that leads to Eq. \eqref{eq:Tvar_GR1_computed} and the final result reads
\begin{align} \label{eq:Tvar_GR2_computed}
{\rm Tvar}^\nu_{2}(K^c_1,K^c_2,K_3^c) =& {\rm Tvar}^\nu_{1}(K^c_1,K^c_2,K_3^c) -2 H^2 \pi \, \frac{K_2^c}{\left(2 K_1^c\right)^{5}} \, \frac{\Gamma(9/2-\nu) \Gamma(9/2+\nu)}{\Gamma(5)} \nonumber \\
&\qquad\qquad\qquad\qquad \times {\rm H2F1}\left(9/2-\nu, \,9/2+\nu, \, 5 , \,\frac{1}{2} - \frac{K_2^c+K_3^c}{2 K_1^c}\right)  \, . 
\end{align}
Again, this expression is valid providing that $\nu \neq 3/2$. In the case $\nu = 3/2$ we get the result
\begin{align} \label{eq:Tvar2}
{\rm Tvar}^{\nu=3/2}_{2}(K^c_1,K^c_2,K_3^c) = 0 \, .
\end{align}  
Another key difference with respect to the scalar-scalar-tensor computation is that here we need to sum over the following two independent sets of momenta and helicities
\begin{align}
\mathbf{K^{c_1}} = \{ \bk_1,\bk_2,\bk_3\} \, , \qquad\qquad \mathbf{K^{c_2}} = \{ \bk_1,\bk_3,\bk_2 \} \, ,
\end{align}
and 
\begin{align}
\lambda^{c_1} = \{0,\lambda_2,\lambda_3\} \, , \qquad\lambda^{c_2} =  \{0,\lambda_3,\lambda_2\} \, .  
\end{align}
In the end we find
\begin{align} \label{eq:stt_bisp}
\langle \zeta(\bk_1) h_{\lambda_2}(\bk_2) h_{\lambda_3}(\bk_3) \rangle &=  -  \alpha_{\lambda_3} \frac{H^4 \kappa_{CS} \kappa_{\sigma}}{M^4_{pl}} \,\frac{1}{k_1 k_2} \nonumber \\
&\qquad \times \Big[ -2\left(e^{\lambda_2}_{ij}(\hat{k}_2) \cdot e^{\lambda_3}_{ij}(\hat{k}_3)\right) {\rm Tvar}^\nu_{1}(k_1,k_2,k_3)\nonumber \\
&\qquad\qquad+\left(e^{\lambda_2}_{ij}(\hat{k}_2) \cdot e^{\lambda_3}_{ij}(\hat{k}_3)\right) {\rm Tvar}^\nu_{2}(k_1,k_2,k_3)\nonumber \\
&\qquad\qquad+ \frac{1}{(k_3)^2} \left((k_2)_l \,(k_3)_m \cdot e^{\lambda_2}_{im}(\hat{k}_2) e^{\lambda_3}_{il}(\hat{k}_3) \right) {\rm Tvar}^\nu_{2}(k_1,k_3,k_2)\nonumber \\
&\qquad\qquad-\frac{(k_2)_l \,(k_3)_l}{(k_3)^2}  \left(e^{\lambda_2}_{ij}(\hat{k}_2) \cdot e^{\lambda_3}_{ij}(\hat{k}_3)\right) {\rm Tvar}^\nu_{2}(k_1,k_3,k_2)  \Big]   \nonumber\\
&\qquad\qquad\qquad\qquad\qquad\qquad\qquad\qquad\qquad\qquad+ (k_2 \leftrightarrow k_3, \, \lambda_2 \leftrightarrow \lambda_3) \, .
\end{align}
Note that we get a non-zero result for either the case $\lambda_2 = \lambda_3$ and $\lambda_2 \neq \lambda_3$, therefore in this theory scalar-tensor-tensor bispectra with mixed helicities are allowed. 

\subsection{Comments on the results}

\begin{figure}
        \centering
        \includegraphics[width=.49\textwidth]{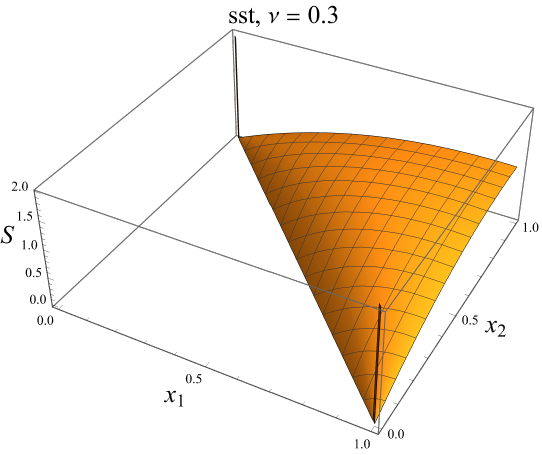} 
       \includegraphics[width=.49\textwidth]{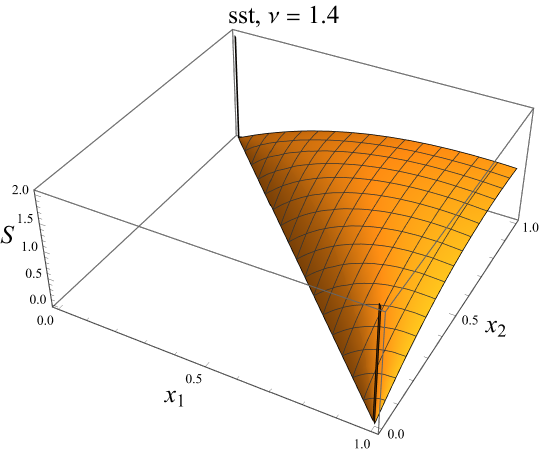}
    \caption{Shape function of the scalar-scalar-tensor bispectrum. The plot is realized as function of $x_1 = k_1/k_3$ and $x_2= k_2/k_3$ and normalized to unity in the equilateral configuration.} \label{fig:shape_sst}
\end{figure}

\begin{figure}
        \centering
        \includegraphics[width=.49\textwidth]{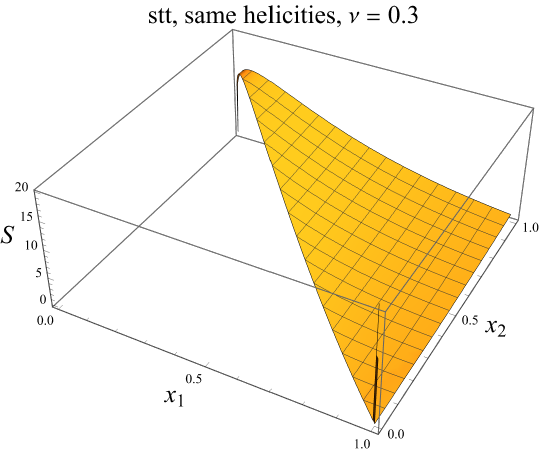} 
       \includegraphics[width=.49\textwidth]{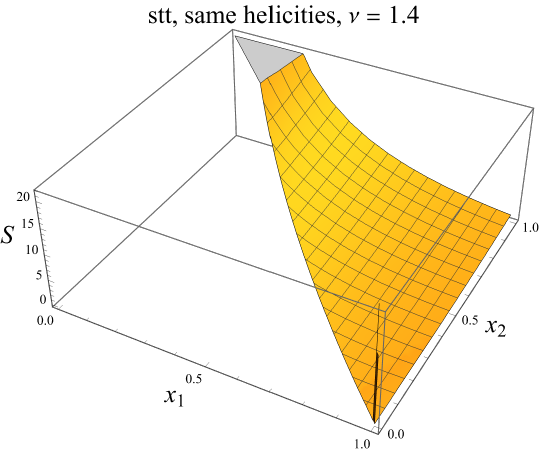}
    \caption{Shape function of the scalar-tensor-tensor bispectrum when tensor perturbations have the same helicity. The plot is realized as function of $x_1 = k_1/k_3$ and $x_2= k_2/k_3$ and normalized to unity in the equilateral configuration.} \label{fig:shape_stt_same}
\end{figure}

\begin{figure}
        \centering
        \includegraphics[width=.49\textwidth]{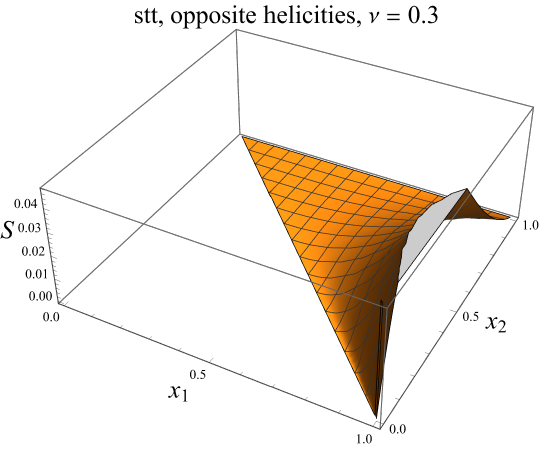} 
       \includegraphics[width=.49\textwidth]{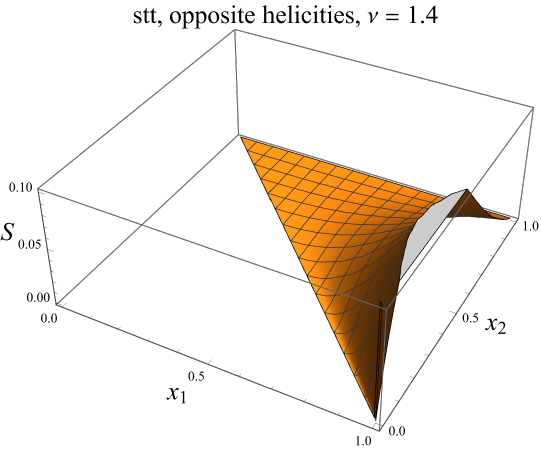}
    \caption{Shape function of the scalar-tensor-tensor bispectrum when tensor perturbations have opposite helicities. The plot is realized as function of $x_1 = k_1/k_3$ and $x_2= k_2/k_3$.} \label{fig:shape_stt_opposite}
\end{figure}

We now comment on the results obtained. First, consider the helicity factor $\alpha_\lambda$ appearing in the expressions for the bispectra in Eqs.~\eqref{eq:sst_bisp} and \eqref{eq:stt_bisp}. This factor implies that the bispectra are parity-odd, as evidenced by the following relations:
\begin{align}
\langle \zeta(\mathbf{k}_1) \zeta(\mathbf{k}_2) h_{R}(\mathbf{k}_3) \rangle &= - \langle \zeta(\mathbf{k}_1) \zeta(\mathbf{k}_2) h_{L}(\mathbf{k}_3) \rangle \,,
\end{align}
for the scalar-scalar-tensor bispectrum, and
\begin{align}
\langle \zeta(\mathbf{k}_1) h_{R}(\mathbf{k}_2) h_{R}(\mathbf{k}_3) \rangle &= - \langle \zeta(\mathbf{k}_1) h_{L}(\mathbf{k}_2) h_{L}(\mathbf{k}_3) \rangle \,, \\
\langle \zeta(\mathbf{k}_1) h_{R}(\mathbf{k}_2) h_{L}(\mathbf{k}_3) \rangle &= - \langle \zeta(\mathbf{k}_1) h_{L}(\mathbf{k}_2) h_{R}(\mathbf{k}_3) \rangle \,,
\end{align}
for the scalar-tensor-tensor bispectra. These relations directly reflect the violation of parity symmetry in the theory.

When the mediating field $\sigma$ is massless ($\nu = 3/2$), these parity-odd bispectra vanish due to the vanishing of the corresponding time-ordered variances, as shown in Eqs.~\eqref{eq:Tvar1} and \eqref{eq:Tvar2}. This result is consistent with general theorems in the literature concerning parity-odd correlators involving massless fields, where in-in time integrals are globally infrared (IR) convergent (see, e.g., \cite{Liu:2019fag,Cabass:2022rhr,Stefanyszyn:2023qov,Thavanesan:2025kyc}). This conclusion holds provided the contribution from the right diagram in Fig.~\ref{fig:sst_diagrams_sigma_med} is neglected, which is justified a priori by Eq.~\eqref{eq:RoverL}. However, since the left diagram yields an exactly vanishing contribution in this case, the right diagram could potentially dominate, provided its time integrals are not globally IR-convergent, e.g., if they exhibit a logarithmic divergence. 

An explicit calculation shows that in the massless limit, the right diagram indeed yields a non-zero result to the scalar-scalar-tensor (sst) bispectrum. This contribution is proportional to $\left(N_{2k_1} + N_{2k_2}\right)^2$, where $N_K = \log(-K \tau_0)$ represents the number of e-folds between horizon exit of the mode with wavenumber $K$ and the end of inflation. Nonetheless, we do not explore this contribution further, as our focus is on scenarios in which the mediating field has a mass comparable to $H$.

When the mediating field $\sigma$ is massive ($\nu < 3/2$), we obtain a non-zero contribution from the left diagram in Fig.~\ref{fig:sst_diagrams_sigma_med}, which in this regime provides the dominant contribution to the sst bispectrum. Also the scalar-tensor-tensor (stt) bispectrum gets a non-zero result. Notably, the Gauss hypergeometric functions appearing in Eqs.~\eqref{eq:Tvar_GR1_computed} and \eqref{eq:Tvar_GR2_computed} exhibit poles when $k_t = k_1 + k_2 + k_3 = 0$. As discussed in \cite{Stefanyszyn:2023qov}, these poles correspond to a so-called fake total-energy singularity, which arises from a partial-energy singularity transformed into a total-energy singularity due to momentum conservation at the quadratic vertex $\delta \sigma \zeta$.

Once the momenta $\mathbf{k}_i$ are fixed in space, the Kronecker-delta contractions involving Latin indices in Eqs.~\eqref{eq:sst_bisp} and \eqref{eq:stt_bisp} can be rewritten in terms of the corresponding wave numbers $k_i$. For this purpose, the simplest and most convenient configuration is the one described in Appendix~\ref{app:pol_ten}, where one of the graviton momenta, $k_3$, is aligned along the $x$-axis.

In Figs.~\ref{fig:shape_sst}, \ref{fig:shape_stt_same}, and \ref{fig:shape_stt_opposite}, we present the shape function of the bispectra, specifically the quantity $k_1^2 k_2^2 k_3^2 \, B(k_1,k_2,k_3)$, plotted as a function of the dimensionless ratios $x_1 = k_1/k_3$ and $x_2 = k_2/k_3$. We show results for two different values of the mass parameter of the spectator field: $\nu = 0.3$, corresponding to a heavier spectator field ($m_{\sigma} > H$), and $\nu = 1.4$, corresponding to a lighter spectator field ($m_{\sigma} < H$), with respect to $H$.

Figure~\ref{fig:shape_sst} shows that the sst bispectrum peaks in the equilateral configuration and vanishes in the squeezed limit.

Figure~\ref{fig:shape_stt_same} illustrates the stt bispectrum in the case where the tensor perturbations have the same helicity. In this case, the bispectrum exhibits a peak in the squeezed isosceles limit ($k_1 \ll k_2 \simeq k_3$) and remains nonzero in the equilateral configuration. However, the scaling behavior in the squeezed limit follows $k_L^{-3/2 - \nu}$, which is milder than that of the standard local non-Gaussianity ($\propto k_L^{-3}$) when $\nu < 3/2$.

Figure~\ref{fig:shape_stt_opposite} depicts the stt bispectrum for the case of tensor perturbations with opposite helicities. In this case, the bispectrum vanishes in both the equilateral and squeezed configurations and reaches its maximum for isosceles configurations with $k_1 \simeq k_3$.

\subsection{Perturbativity bound: strength of non-Gaussianity} \label{sec:pert_bound}

\begin{figure}
        \centering
        \includegraphics[width=.7\textwidth]{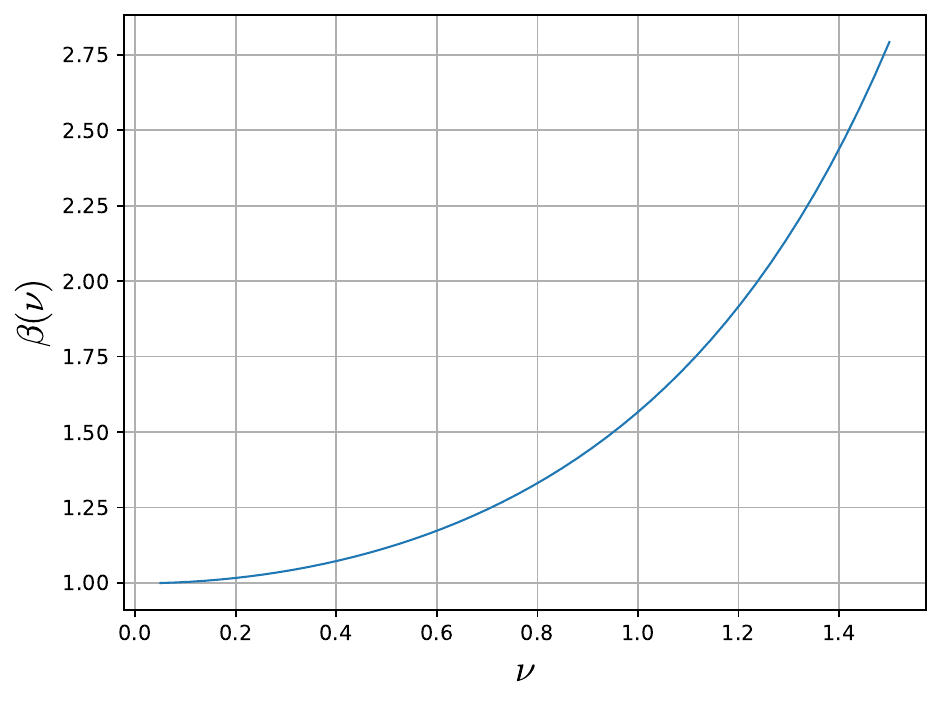} 
    \caption{Parameter $\beta(\nu)$ that appears in Eq. \eqref{eq:bispectrum_equi}.}  \label{fig:bisp_vs_nu}
\end{figure}

Normally, we define how much primordial bispectra lead to the departure from a Gaussian behaviour in primordial perturbations by taking the ratio between the bispectrum evaluated in the equilateral configuration and the product of two scalar power spectra. This in formula reads
\begin{align}  \label{eq:fNL_def}
f_{\rm NL} = \frac{B(k,k,k)}{P^2_{\zeta}(k)} \, ,
\end{align}
where
\begin{align}
P_{\zeta}(k) = \frac{H^2}{M^2_{pl} 4 \epsilon \, k^3} \, .
\end{align}
As a reference in the standard slow-roll model of inflation we have \cite{Maldacena:2002vr}
\begin{align}
f^{sst}_{\rm NL} \sim \epsilon \, .
\end{align}
As claimed e.g. in \cite{Cabass:2021fnw}, if primordial scalar-tensor non Gaussianities are non-zero and can be detected, $\langle \zeta \zeta h \rangle$ is the cosmological correlator which allows to make a detection with the highest signal-to-noise ratio. Therefore, an observability forecast should preferably target this sst bispectrum. Evaluating the sst bispectrum in Eq. \eqref{eq:sst_bisp} in the equilateral configuration we have
\begin{align}\label{eq:bispectrum_equi}
|B^{\zeta\zeta h}_{PV}(k,k,k)| k^6 \simeq  0.22 \, \beta(\nu) \, \frac{H^4}{M^4_{pl}} (H \kappa_{CS}) \, (H \kappa_{\sigma})  \, ,
\end{align}
where $\beta(\nu)$ is plotted in Fig. \ref{fig:bisp_vs_nu}. 
It follows that the strength of non-Gaussianity can be quantified as
\begin{align} \label{eq:fNL_equi}
f^{sst, PV}_{\rm NL} \simeq 3.52 \, \beta(\nu) \, \epsilon^2 (H\kappa_{CS}) \, (H \kappa_{\sigma})  \, .
\end{align}
Despite the apparent large slow-roll suppression of this result, in order to quantify the upper bound to this amplitude we shall determine the maximum strength of the couplings $\kappa_{CS}$ and $\kappa_{\sigma}$ compatible with our effective field theory.

\begin{figure}
        \centering
        \includegraphics[width=.49\textwidth]{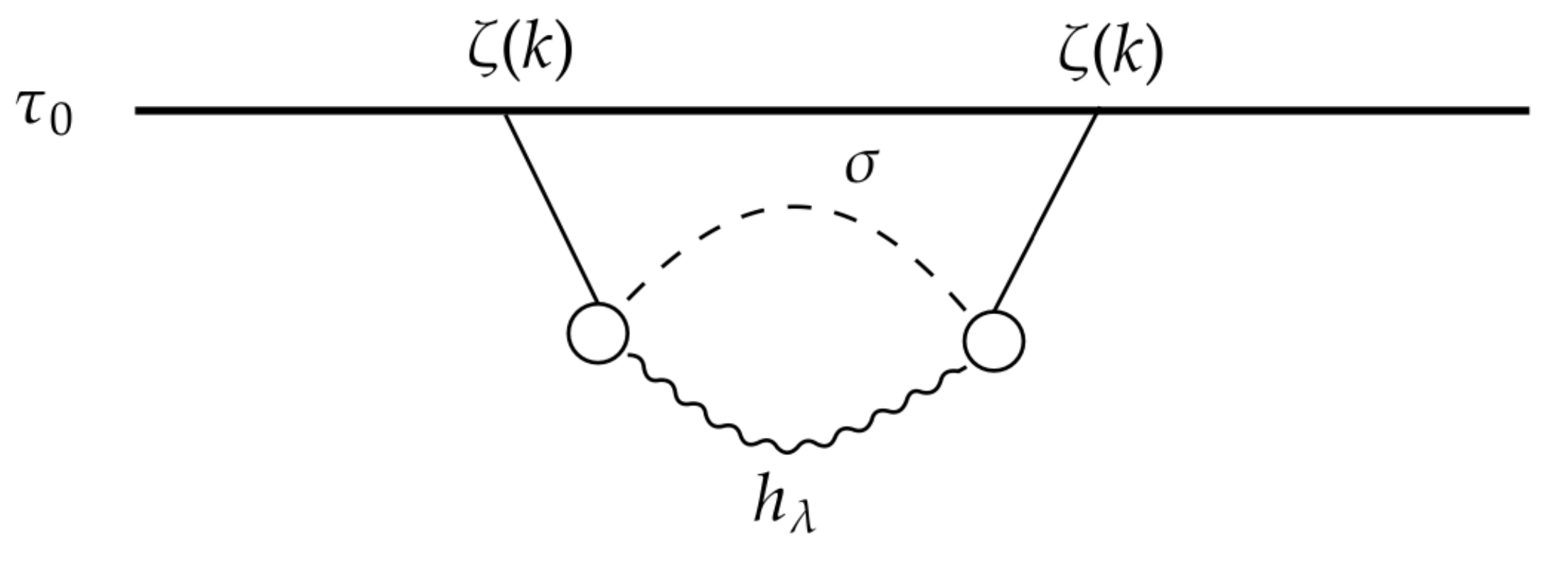} 
    
    \caption{Qualitative Feynman diagram for the 1-loop scalar power spectrum correction from $\delta \sigma 
    \zeta h$ vertices.} \label{fig:sst_diagram_radiative}
\end{figure}

\begin{figure}
        \centering
        \includegraphics[width=.7\textwidth]{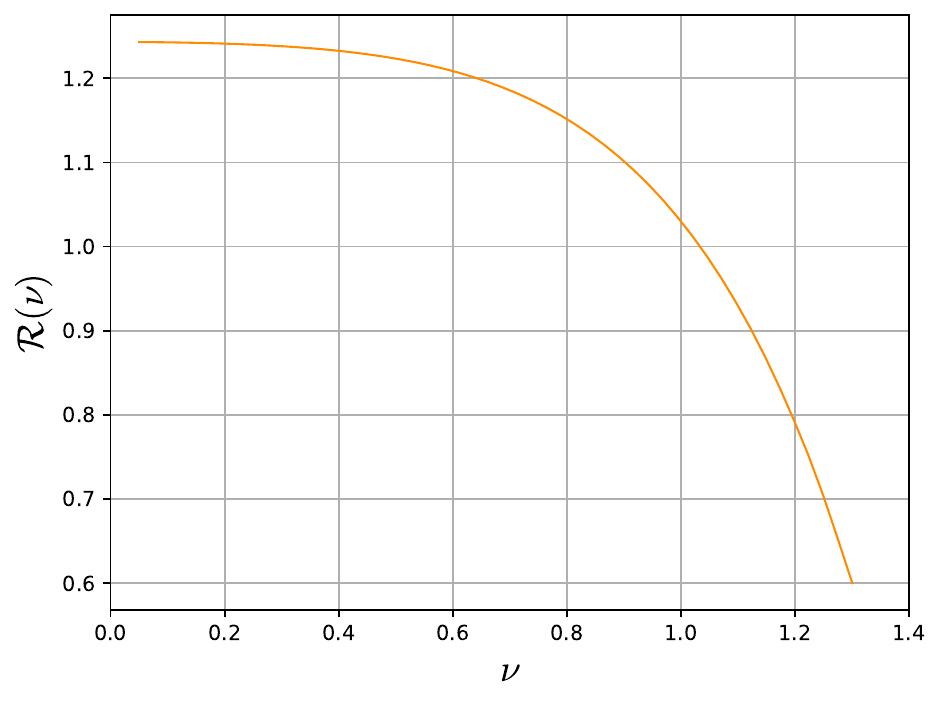} 
    \caption{Parameter $\mathcal R(\nu)$ as in Eq. \eqref{eq:Rnu}.}  \label{fig:Rnu}
\end{figure}

The maximum strength of the coupling $\kappa_{CS}$ before reaching the strong coupling regime comes from the computation of the 1-loop correction to the scalar power spectrum provided by the vertex $\delta\sigma\zeta h$. Here we perform only an order-of-magnitude estimation (up to factors of order one). The Feynman diagram corresponding to this computation is given in Fig. \ref{fig:sst_diagram_radiative}. This diagram gives a corrective contribution to the power spectrum of the order
\begin{align}
\frac{\delta P_\zeta}{P_\zeta} \sim \frac{H^4}{M^4_{pl}} \epsilon \, (H \kappa_{CS})^2 \simeq 10^{-16} \epsilon^3 \, (H \kappa_{CS})^2\, . 
\end{align}
It follows that in order to ensure that radiative corrections are under control we must impose 
\begin{align} \label{eq:condition_coupling1}
\kappa_{CS} \ll \frac{10^{8} \epsilon^{-3/2}}{H} \, . 
\end{align}
The other coupling constant $\kappa_{\sigma}$ is constrained by the condition \eqref{eq:radiative_control_mass_precise}, which is placed in order to prevent significant tree level corrections to the scalar power spectrum. It follows that the product between the coupling constants is constrained to be 
\begin{align} \label{eq:constraint_couplings}
\kappa_{CS} \kappa_{\sigma} \ll \frac{10^{8} \epsilon^{-2}}{2HM_{pl}\sqrt{\mathcal C(\nu)}} \, .
\end{align}
By applying the upper bound from Eq.~\eqref{eq:constraint_couplings} along with the relation in Eq.~\eqref{eq:HvsMpl} to Eq.~\eqref{eq:fNL_equi}, we obtain the following upper bound on the coefficient of non-Gaussianity in terms of the slow-roll parameter:
\begin{align}
f^{\text{sst, PV}}_{\rm NL} \ll 1.76 \times 10^{4} \, \mathcal R(\nu) \, \sqrt{\epsilon} \,,
\end{align}
where $\mathcal R(\nu)$ denotes the ratio
\begin{align} \label{eq:Rnu}
\mathcal R(\nu) = \beta(\nu)/\sqrt{\mathcal{C}(\nu)} \, .
\end{align}
We plot this quantity in Fig. \ref{fig:Rnu}. Taking into account the current observational upper limit on the tensor-to-scalar ratio, $r \lesssim 0.01$, and using the standard slow-roll relation $r = 16 \epsilon$, we arrive at the following numerical bound:
\begin{align} \label{eq:est_fNL}
f^{\text{sst, PV}}_{\rm NL} \ll 440 \,\mathcal R(\nu) \,,
\end{align}
which depends on the mass parameter $\nu$. From Fig. \ref{fig:Rnu} we see that in the regime where the spectator field has a relatively large mass ($\nu \lesssim 0.4$), the ratio $\mathcal R(\nu) \simeq 1.24$, yielding an approximately mass-independent constraint:
\begin{align} \label{eq:est_fNL_massive}
f^{\text{sst, PV}}_{\rm NL} \big|_{\nu \lesssim 0.4} \ll 546 \,.
\end{align}
In contrast, when the spectator field is lighter ($\nu > 0.4$), the function $\mathcal{R}(\nu)$ decreases monotonically, thereby gradually reducing the potential observability of the signal. Nevertheless, even for lighter masses, like $\nu = 1.3$, the constraint reads
\begin{align} \label{eq:est_fNL_massive2}
f^{\text{sst, PV}}_{\rm NL} \big|_{\nu =1.3} \ll 264 \, ,
\end{align}
which is of the same order of magnitude as \eqref{eq:est_fNL_massive}. Only for tiny masses the constraint on $f^{\text{sst, PV}}_{\rm NL}$ loses orders of magnitude with respect to \eqref{eq:est_fNL_massive}. For example when $\nu = 1.45$ we obtain
\begin{align}
\label{eq:est_fNL_massive3}
f^{\text{sst, PV}}_{\rm NL} \big|_{\nu =1.45} \ll 36 \, .
\end{align}
However, so far we have not considered the constraint in Eq. \eqref{eq:conditions_sigmadotCS2}:
\begin{align} 
\kappa_{CS} \kappa_\sigma \ll \frac{m_\sigma^2}{H^2}\frac{10^4\epsilon^{-3/2} |\epsilon-\eta|^{-1}}{16 H M_{pl}} \, .
\end{align}
Even taking the maximum value for the mass of the spectator field, $m_\sigma = 3/2 H$, we get
\begin{align} \label{eq:conditions_sigmadotCS4}
\kappa_{CS} \kappa_\sigma \ll 1.4 
\frac{10^3 \epsilon^{-3/2} |\epsilon-\eta|^{-1}}{H M_{pl}}  \, .
\end{align}
This constraint is a factor $10^5\epsilon^{-1/2} |\epsilon-\eta|$ stronger than that in Eq. \eqref{eq:constraint_couplings}. For $r \lesssim 0.01$ as above, $\epsilon \sim 10^{-3}$, $\eta \sim 10^{-2}$ due to the scalar tilt relation of slow-roll models of inflation, and we get the tighter constraint
\begin{align} \label{eq:conditions_sigmadotCS3}
f^{\text{sst, PV}}_{\rm NL} \ll 2 \times 10^{-2} \, .
\end{align}
While this constraint is better than that generated by the traditional Chern-Simons term with the inflaton\footnote{Using the same normalization of this work, the  coefficient $f^{\text{sst, PV}}_{\rm NL}$ induced by the traditional gCS-term coupled to the inflaton field is of the order \cite{Bartolo:2017szm}
\begin{align} 
f^{\text{sst, PV}}_{\rm NL}|_{\phi-CS} \sim \epsilon \frac{H}{M_{CS}} \ll 10^{-3} 
\end{align}
when $\epsilon \sim 10^{-3}$.}
, its strength still make a detection challenging. In fact, as suggested by the forecasts in \cite{Bartolo:2018elp,Shiraishi:2019yux}, scalar-tensor non-Gaussianities normalized as in our work are detectable by CMB experiments targeting the CMB polarization field if at least $f_{\rm NL} \sim \mathcal O(1)$. Therefore, while perturbative considerations in principle allow for sizable and observable effects on primordial non-Gaussianity, the dynamics of $\sigma_0$ enforces the attenuation of the signatures to a level where a detection is difficult with the instruments we have so far. This suggests that one should search for an alternative transfer mechanism where the $\sigma$-background dynamics is completely frozen, i.e. $\dot \sigma_0 = 0$, or at least where $\dot \sigma_0$ is allowed to be smaller with respect to the present model. 


\section{Summary and conclusion}
\label{sec:conclusion}
In this work, we have investigated the gravitational Chern-Simons (gCS) operator minimally coupled to a spectator field during multi-field inflation. As a minimal two-field toy model, we considered a setup where one field $\phi$ plays the role of the standard slowly-rolling inflaton, while the second field $\sigma$ is a massive spectator field in a regime where it contributes negligibly to the dynamics of inflation. Also, we introduced a kinetic coupling between the two fields preserving the approximate shift symmetry of the inflaton. As a result a linear interaction term emerges between the perturbations of the two fields. In addition, a centrifugal force appears in the background equation of motion for $\sigma$ which could potentially spoil the whole background dynamics. To avoid this, we assumed that this centrifugal force is compensated by the spectator field self-potential and derived appropriate bounds on the kinetic coupling constant that keep the effects of the spectator field on the inflationary background and the scalar power spectrum negligible. 

In this scenario, the spectator field contributes perturbations, which—if sufficiently massive—decay before the end of inflation and leave no direct observable imprint. However, due to the linear $\delta\sigma\zeta$ coupling, indirect imprints can be transferred to correlators of curvature perturbations, offering a window into the inflationary particle content that has decayed. 

We found that while the usual ghost instability of Chern–Simons gravity constrains the product of the time derivative of the spectator field and the gCS–$\sigma$ coupling constant, i.e. $\kappa_{CS}\dot\sigma_0$, the leading cubic interactions depend solely on $\kappa_{CS}$. Consequently, if $\dot\sigma_0$ is sufficiently small, this feature in principle relaxes the constraint on $\kappa_{CS}$, thereby enhancing the maximum strength of non-linear interactions relative to the standard scenario in which the gCS term is coupled to the inflaton. We explicitly identified novel parity-violating interactions of the form $\delta\sigma h h$, $\delta\sigma\zeta h$, and $\delta\sigma\delta\sigma h$ (Eqs.~\eqref{eq:hhsigma}–\eqref{eq:hsigmasigma}). These couplings in combination with the exchange vertex $\delta\sigma\zeta$ generate distinctive parity-odd scalar-tensor bispectra $\langle \zeta\zeta h \rangle$ and $\langle \zeta h h \rangle$, which we computed using the Schwinger-Keldysh formalism. The contributions were evaluated analytically for arbitrary spectator field mass, even if the focus of this work is for intermediate masses $m_\sigma \sim H$.

Following perturbative considerations, we derived theoretical upper limits on the product of the $\delta\sigma$-gCS and $\delta\sigma\zeta$ coupling constants that enter the bispectra amplitude (Eq. \eqref{eq:constraint_couplings}). Implementing the current upper bound on $r$, the resulting constraint on the parity-odd scalar-scalar-tensor non-Gaussianity parameter, $f^{\text{sst, PV}}_{\rm NL}$, allows for values of order unity (Eq. \eqref{eq:est_fNL} together with Fig. \ref{fig:Rnu}). However, when considering the consistency constraint due to the avoidance of the ghost-field regime, the theoretical upper bound becomes stronger, leading in the best case scenario to $f^{\text{sst, PV}}_{\rm NL} \ll 2 \times 10^{-2}$ for $r$ of the order of the current bound (Eq. \eqref{eq:conditions_sigmadotCS3} and text before). While this represents an improvement with respect to the usual Chern-Simons gravity with the inflaton field, indeed it makes a detection challenging.

We want to stress that modifications to our model including also different transfer mechanisms might result in a modeling where $\dot\sigma_0$ is allowed to be smaller than in this work or equal to $0$, allowing for sizable signatures in non-Gaussianities and/or alternative shape functions. 


For what summarized, our findings motivate further exploration of parity-violating signatures from spectator-gCS interactions during inflation. In particular, it would be valuable to systematically investigate alternative coupling structures between curvature and spectator modes or other mechanisms of signature transfer. Furthermore, similar effects could emerge in the scalar trispectrum, offering an additional probe of parity violation. Detectability forecasts for the predicted bispectra in upcoming CMB experiments will also be essential. We leave these and related questions for future work.

\section*{Acknowledgements} 
The work of G.O. was supported by grant No. UMO-2021/42/E/ST9/00260 from the National Science Center, Poland.

\appendix

\section{Computation of time integrals} \label{app:time_var_comp}

In this appendix we show how to compute analytically the integrals of the kind in Eqs. \eqref{I1_comp} and \eqref{I2_comp} for a generic value of the mass-parameter $0\leq\nu\le3/2$. First, we write down the integrals to be computed
\begin{align}
I^{(1)}_{1}(K^c_1,K^c_2,K_3^c) =& \left(\prod_i (1 + i K^c_i \tau_0) \, e^{- i K^c_i \tau_0} \right) \nonumber \\
&\times \Big[ \int_{-\infty}^{\tau_0}  d \tau_1 \, \tau_1 \, v_\sigma (\tau_1, K_1^c) \, e^{i (K_2^c+K_3^c) \tau_1} \int_{-\infty}^{\tau_1}  \,  \frac{d \tau_2}{\tau^2_2} \, v^{*}_\sigma (\tau_2, K_1^c) \, e^{i K_1^c \tau_2}    \nonumber \\
& \qquad\qquad + \, \int_{-\infty}^{\tau_0}  d \tau_1 \, \tau_1  \, v^*_\sigma (\tau_1, K_1^c) \, e^{i (K_2^c + K_3^c) \tau_1} \int_{\tau_1}^{\tau_0} \frac{d \tau_2}{\tau^2_2} \,v_\sigma (\tau_2, K_1^c) \, e^{i K_1^c \tau_2} \Big] \, , 
\end{align}  
\begin{align} 
I^{(2)}_{1}(K^c_1,K^c_2,K_3^c) =& -(1 + i K^c_1 \tau_0) \, (1 - i K^c_2 \tau_0) \, (1 - i K^c_3 \tau_0) \, e^{- i (K^c_1-K^c_2-K^c_3) \tau_0} \nonumber \\
& \times \Big[ \int_{-\infty}^{\tau_0}  d \tau_1 \, \tau_1 \, v_\sigma (\tau_1, K_1^c)  \, e^{-i (K_2^c+K_3^c) \tau_1}  \int_{-\infty}^{\tau_0}  \frac{d \tau_2}{\tau^2_2} \, v^{*}_\sigma (\tau_2, K_1^c) \, e^{i K_1^c \tau_2} \Big]  \, . 
\end{align}
Given the IR-behavior of the mode function $v_\sigma \propto (-\tau_0)^{3/2-\nu}$ (from Eq. \eqref{eq:asympthotic_sigma}), we get the following scaling behaviour of the integrals in terms of $\tau_0$  
\begin{align} 
I^{(j)}_{1}&\sim \left[1+\mathcal O(\tau_0^2)\right] (-\tau_0)^{1/2-\nu} + \mbox{\, FINITE $(\tau_0 \rightarrow 0)$} \, ,
\end{align}  
where the contribution inside the square parenthesis denotes the order in $\tau_0$ provided by the $\tau_0$-dependent coefficients in front of the integrals. It follows that the integrals contain some IR-divergences for $\nu>1/2$. However, as we will show later on, these divergences are spurious and are not contained in the final result, which is finite. Also, since the $\tau_0$-dependent coefficients provide contributions equal to the unity in the limit $\tau_0 \rightarrow 0$, we can drop them and re-write the previous integrals as
\begin{align}
I^{(1)}_{1}(K^c_1,K^c_2,K_3^c) =& \Big[ \int_{-\infty}^{\tau_0}  d \tau_1 \, \tau_1 \, v_\sigma (\tau_1, K_1^c) \, e^{i (K_2^c+K_3^c) \tau_1} \int_{-\infty}^{\tau_1}  \,  \frac{d \tau_2}{\tau^2_2} \, v^{*}_\sigma (\tau_2, K_1^c) \, e^{i K_1^c \tau_2}    \nonumber \\
& \qquad\qquad - \, \int_{-\infty}^{\tau_0}  d \tau_1 \, \tau_1  \, v^*_\sigma (\tau_1, K_1^c) \, e^{i (K_2^c + K_3^c) \tau_1} \int_{-\infty}^{\tau_1} \frac{d \tau_2}{\tau^2_2} \,v_\sigma (\tau_2, K_1^c) \, e^{i K_1^c \tau_2} \nonumber \\
& \qquad\qquad + \, \int_{-\infty}^{\tau_0}  d \tau_1 \, \tau_1  \, v^*_\sigma (\tau_1, K_1^c) \, e^{i (K_2^c + K_3^c) \tau_1} \int_{-\infty}^{\tau_0} \frac{d \tau_2}{\tau^2_2} \,v_\sigma (\tau_2, K_1^c) \, e^{i K_1^c \tau_2} \Big] \, , 
\end{align}  
\begin{align} 
I^{(2)}_{1}(K^c_1,K^c_2,K_3^c) =& -\Big[ \int_{-\infty}^{\tau_0}  d \tau_1 \, \tau_1 \, v_\sigma (\tau_1, K_1^c)  \, e^{-i (K_2^c+K_3^c) \tau_1}  \int_{-\infty}^{\tau_0}  \frac{d \tau_2}{\tau^2_2} \, v^{*}_\sigma (\tau_2, K_1^c) \, e^{i K_1^c \tau_2} \Big]  \, . 
\end{align}
As in the end we need to take the sum between these two integrals, it is convenient to re-define them in a way to separate the nested integrations from the factorized contributions. Therefore, we re-write the integrals as
\begin{align}
\tilde I^{(1)}_{1}(K^c_1,K^c_2,K_3^c) =& \int_{-\infty}^{\tau_0}  d \tau_1 \, \tau_1  \, e^{i (K_2^c+K_3^c) \tau_1} \int_{-\infty}^{\tau_1}  \,  \frac{d \tau_2}{\tau^2_2} \, e^{i K_1^c \tau_2}\, \Big[v_\sigma (\tau_1, K_1^c) \, v^{*}_\sigma (\tau_2, K_1^c)  - c.c.\Big]  \, , 
\end{align}  
\begin{align} 
\tilde I^{(2)}_{1}(K^c_1,K^c_2,K_3^c) =&  \int_{-\infty}^{\tau_0}  \frac{d \tau_2}{\tau^2_2}  \, e^{i K_1^c \tau_2} \int_{-\infty}^{\tau_0}  d \tau_1 \, \tau_1 \, \Big[v^*_\sigma (\tau_1, K_1^c)  v_\sigma (\tau_2, K_1^c) \, e^{i (K_2^c+K_3^c) \tau_1} - c.c. \Big]  \, , 
\end{align}
where $\tilde I_{1}^{(1)}+\tilde I_{1}^{(2)} = I_{1}^{(1)}+I_{1}^{(2)}$. By using the decomposition of the Hankel functions in terms of the Bessel functions of the first and second kind
\begin{align} 
H_\nu^{(1)}(- k \tau) = J_\nu(- k \tau) + i \, Y_{\nu}(- k \tau) \, , \\
H_\nu^{(2)}(- k \tau) = J_\nu(- k \tau) - i \, Y_{\nu}(- k \tau) \, , 
\end{align}  
and exploiting the fact that the Bessel functions are purely real for real positive arguments, we can decompose the integrations above as
\begin{align} \label{eq:I1tilde}
\tilde I^{(1)}_{1} =& i \frac{\pi}{2} H^2 \Big[\int_{-\infty}^{\tau_0}  d \tau_1 \, (-\tau_1)^{5/2}  \,J_\nu(- K_1^c \tau_1)\, e^{i (K_2^c+K_3^c) \tau_1} \int_{-\infty}^{\tau_1}  \, d \tau_2 \, (-\tau_2)^{-1/2} \,Y_\nu(- K_1^c \tau_2)\, e^{i K_1^c \tau_2}\, \nonumber \\
& -\int_{-\infty}^{\tau_0}  d \tau_1 \, (-\tau_1)^{5/2}  \,Y_\nu(- K_1^c \tau_1) \, e^{i (K_2^c+K_3^c) \tau_1} \int_{-\infty}^{\tau_1}  \,  d \tau_2\,(-\tau_2)^{-1/2} J_\nu(- K_1^c \tau_2)\,e^{i K_1^c \tau_2}\, \Big]\, , 
\end{align}  
and
\begin{align}  \label{eq:I2tilde_decom}
\tilde I^{(2)}_{1}=&  \frac{\pi}{2} H^2\Big\{{\rm Re}\Big[\int_{-\infty}^{\tau_0}  d \tau_1 \, (-\tau_1)^{5/2} \, H^{(2)}_\nu (-K_1^c \tau_1) \, e^{i (K_2^c+K_3^c)\tau_1} \Big] {\rm Im}\Big[\int_{-\infty}^{\tau_0}  \,  d \tau_2 \,(-\tau_2)^{-1/2}\, Y_\nu(-K_1^c \tau_2)  \, e^{i K_1^c \tau_2}\Big]  \nonumber\\
& +{\rm Im}\Big[\int_{-\infty}^{\tau_0}  d \tau_1 \, (-\tau_1)^{5/2} \, H^{(2)}_\nu (-K_1^c \tau_1) \, e^{i (K_2^c+K_3^c)\tau_1}   \Big] {\rm Im}\Big[\int_{-\infty}^{\tau_0}  \,  d \tau_2 \,(-\tau_2)^{-1/2}\, J_\nu(-K_1^c \tau_2)  \, e^{i K_1^c \tau_2}\Big]\Big\} \nonumber\\
& \qquad\qquad\qquad\qquad +i \Big(...\Big)\, , 
\end{align}
where the $+i \Big(...\Big)$ denotes a pure imaginary contribution which will not contribute as in Eq. \eqref{eq:Tvar_GR1} we need to take the real part of the final result.

The Bessel function of the second kind can be expressed in terms of the Bessel function of the first kind as
\begin{align} \label{eq:YvsJ}
Y_{\nu}(x) = \frac{J_{\nu}(x)\cos(\nu \pi) - J_{-\nu}(x)}{\sin (\nu \pi)} \, .
\end{align}
Moreover, we can employ the following series decomposition of the Bessel function of the first kind
\begin{align} \label{eq:Jseries}
J_{\nu}(x) = \frac{\Gamma(2\nu+1)}{\Gamma(\nu+1)\Gamma(\nu+1/2)} \left(\frac{x}{2}\right)^\nu e^{- i x}\sum_{j=0}^{\infty} \frac{\Gamma(\nu+1/2+j)}{\Gamma(2\nu+1+j)\Gamma(j+1)} \,(2 i x)^j\,  ,
\end{align}
where $\Gamma(n)$ denotes the Gamma function. This series expansion reduces integrals of the Bessel functions to integrals of the kind
\begin{align} \label{eq:master_integral}
I_{d}(k,q,\bar\tau) = \int_{-\infty}^{\bar\tau}  d \tau \, (-k \tau)^{d} \, \, e^{i q\tau} = -\,i^{-d+1} \left(\frac{k}{q}\right)^d \frac{1}{q} \, \Gamma(1+d, - i q \bar\tau)  \, ,
\end{align}
which can be expressed in terms of the incomplete Gamma function $\Gamma(n,z)$. In turn, the latter can also be expanded in series as 
\begin{align}\label{eq:Gamma_series}
\Gamma(s,z) = \Gamma(s) - z^s e^{-z} \,\sum_{j=0}^{\infty} \frac{\Gamma(s)}{\Gamma(s+j+1)} \, z^j \, .
\end{align}
This allows to keep performing the nested integrals in terms of Eq. \eqref{eq:master_integral}. However, note that  when $\nu = 1/2, \,1, \,3/2$ the expansion in series of the Bessel function of the second kind in terms of $J_{\nu}$ and $J_{-\nu}$ contains some indeterminate expressions. For the values $\nu = 1/2, \,1$ we can Taylor expand the final result near $\nu = 1/2, \,1$ in order to remove the indeterminate expressions, while for $\nu = 3/2$ we need to perform an independent computation. We will return later on this point. 

Performing the two nested time integrations in Eq. \eqref{eq:I1tilde} as prescribed\footnote{The time integrations have been performed with the help of the software \textit{Mathematica} \cite{Mathematica}.} and taking the limit $\tau_0 \rightarrow 0$, we find 
\begin{align} \label{eq:aux1}
&\int_{-\infty}^{0}  d \tau_1 \, (-\tau_1)^{5/2}  \,J_\nu(- K_1^c \tau_1)\, e^{i (K_2^c+K_3^c) \tau_1} \int_{-\infty}^{\tau_1}  \, d \tau_2 \, (-\tau_2)^{-1/2} \,Y_\nu(- K_1^c \tau_2)\, e^{i K_1^c \tau_2}= \nonumber \\
& = \sum_{j_1,\,j_2,\,j_3=0}^{\infty}\frac{2^{-1 + j_3} \csc(\nu \pi) \,\Gamma(1/2 + j_3 + \nu) \, \Gamma(1 + 2 \nu)}{\sqrt{\pi}
  \,\Gamma(1 + j_1)\,\Gamma(1 + j_3)\,\Gamma(1/2 + \nu)\,\Gamma(1 + \nu)\,\Gamma(1 + j_3 + 2 \nu)} \nonumber \\
&  \qquad \times \Big(-\frac{2^{1/2 - \nu} \,\left(\frac{k_1 + k_2 + k_3}{k_1}\right)^{1/2 - j_3 - \nu}\,
    \Gamma^2(1/2 + j_1 - \nu) \,\Gamma(7/2 + j_3 + \nu)}{(k_1 + k_2 + k_3)^4 \,\Gamma(1 + j_1 + 2 \nu)} \nonumber\\
&  \qquad\qquad + \frac{2^{-1/2 - \nu} \,\left(\frac{k_1 + k_2 + k_3}{k_1}\right)^{1/2 - j_3 - \nu}\,
   \Gamma^2(1/2 + j_1 + \nu) \,\Gamma(1/2 + j_3 + \nu) \,\Gamma(7/2 + j_3 + \nu)}{(k_1 + k_2 + k_3)^4 \,\Gamma(1 + j_3 + 2 \nu)}   \nonumber\\
&  \qquad\qquad + \frac{2^{1 + j_1 + j_2 - 2 \nu} \, \left(\frac{3 k_1 + k_2 + k_3}{k_1}\right)^{-j_1 - j_2 - j_3} \,\Gamma(4 + j_1 + j_2 + j_3) \,\Gamma(1/2 + j_1 - \nu)}{(3 k_1 + k_2 + k_3)^4 \,\Gamma(1 + j_1 - 2 \nu) \,{\rm Pocho}[1/2 + j_1 - \nu,1 + j_2]}    \nonumber\\ 
&  \qquad\qquad - \frac{2^{j_1 + j_2} \,\left(\frac{3 k_1 + k_2 + k_3}{k_1}\right)^{-j_1 - j_2 - j_3 - 2 \nu}\,\Gamma(1/2 + j_1 + \nu) \,\Gamma(4 + j_1 + j_2 + j_3 + 2 \nu)}{(3 k_1 + k_2 + k_3)^4 \,\Gamma(1 + j_1 + 2 \nu) \,{\rm Pocho}[1/2 + j_1 + \nu, 1 + j_2]} \nonumber\\ 
&  \qquad\qquad + \frac{2^{-1/2 - \nu} \,e^{-2 i \nu \pi} \,\left(\frac{k_1 + k_2 + k_3}{k_1}\right)^{1/2 - j_3 - \nu}\, \Gamma(1/2 + j_1 + \nu)^2 \, \Gamma(7/2 + j_3 + \nu)}{(k_1 + k_2 + k_3)^4 \,\Gamma(1 + j_1 + 2 \nu)} \nonumber\\ 
&  \qquad\qquad - \frac{
 2^{j_1 + j_2} \, e^{-2 i \nu \pi} \,\left(\frac{3 k1 + k2 + k3}{k1}\right)^{-j_1 - j_2 - j_3 - 2 \nu}\, \Gamma(1/2 + j_1 + \nu) \,\Gamma(4 + j_1 + j_2 + j_3 + 2 \nu)}{(3 k_1 + k_2 + k_3)^4 \,\Gamma(1 + j_1 + 2 \nu)\, {\rm Pocho}[1/2 + j_1 + \nu, 1 + j_2]} \Big) \, ,
\end{align}
and
\begin{align} \label{eq:aux2}
&\int_{-\infty}^{0}  d \tau_1 \, (-\tau_1)^{5/2}  \,Y_\nu(- K_1^c \tau_1) \, e^{i (K_2^c+K_3^c) \tau_1} \int_{-\infty}^{\tau_1}  \,  d \tau_2\,(-\tau_2)^{-1/2} J_\nu(- K_1^c \tau_2)\,e^{i K_1^c \tau_2} =\nonumber \\
& = \sum_{j_1,\,j_2,\,j_3=0}^{\infty}\frac{2^{-1 + j_3} \csc(\nu \pi) \,\Gamma(1/2 + j_1 + \nu) \, \Gamma(1 + 2 \nu)}{\sqrt{\pi}
  \,\Gamma(1 + j_1)\,\Gamma(1 + j_3)\,\Gamma(1/2 + \nu)\,\Gamma(1 + \nu)\,\Gamma(1 + j_1 + 2 \nu)} \nonumber \\
&  \qquad \times \Big(-\frac{2^{1/2 - 3 \nu} \,\left(\frac{k_1 + k_2 + k_3}{k_1}\right)^{1/2 - j_3 + \nu}\,
    \Gamma(1/2 + j_3 - \nu) \,\Gamma(7/2 + j_3 - \nu) \,\Gamma(1/2 + j_1 + \nu)}{(k_1 + k_2 + k_3)^4 \,\Gamma(1 + j_3 - 2 \nu)} \nonumber\\
&  \qquad\qquad + \frac{2^{-1/2 - \nu} \,\left(\frac{k_1 + k_2 + k_3}{k_1}\right)^{1/2 - j_3 - \nu}\,
   \Gamma(1/2 + j_1 + \nu) \,\Gamma(1/2 + j_3 + \nu) \,\Gamma(7/2 + j_3 + \nu)}{(k_1 + k_2 + k_3)^4 \,\Gamma(1 + j_3 + 2 \nu)}   \nonumber\\
&  \qquad\qquad + \frac{2^{1 + j_1 + j_2 - 2 \nu} \, \left(\frac{3 k_1 + k_2 + k_3}{k_1}\right)^{-j_1 - j_2 - j_3} \,\Gamma(4 + j_1 + j_2 + j_3) \,\Gamma(1/2 + j_3 - \nu)}{(3 k_1 + k_2 + k_3)^4 \,\Gamma(1 + j_3 - 2 \nu) \,{\rm Pocho}[1/2 + j_1 + \nu,1 + j_2]}    \nonumber\\ 
&  \qquad\qquad - \frac{2^{j_1 + j_2} \,\left(\frac{3 k_1 + k_2 + k_3}{k_1}\right)^{-j_1 - j_2 - j_3 - 2 \nu}\,\Gamma(1/2 + j_3 + \nu) \,\Gamma(4 + j_1 + j_2 + j_3 + 2 \nu)}{(3 k_1 + k_2 + k_3)^4 \,\Gamma(1 + j_3 + 2 \nu) \,{\rm Pocho}[1/2 + j_1 + \nu, 1 + j_2]} \nonumber\\ 
&  \qquad\qquad + \frac{2^{-1/2 - \nu} \,e^{-2 i \nu \pi} \,\left(\frac{k_1 + k_2 + k_3}{k_1}\right)^{1/2 - j_3 - \nu}\, \Gamma(1/2 + j_1 + \nu) \,\Gamma(1/2 + j_3 + \nu)\, \Gamma(7/2 + j_3 + \nu)}{(k_1 + k_2 + k_3)^4 \,\Gamma(1 + j_3 + 2 \nu)} \nonumber\\ 
&  \qquad\qquad - \frac{
 2^{j_1 + j_2} \, e^{-2 i \nu \pi} \,\left(\frac{3 k1 + k2 + k3}{k1}\right)^{-j_1 - j_2 - j_3 - 2 \nu}\, \Gamma(1/2 + j_3 + \nu) \,\Gamma(4 + j_1 + j_2 + j_3 + 2 \nu)}{(3 k_1 + k_2 + k_3)^4 \,\Gamma(1 + j_3 + 2 \nu)\, {\rm Pocho}[1/2 + j_1 + \nu, 1 + j_2]} \Big) \, ,
\end{align}
where ${\rm Pocho}[x, n]$ denotes the Pochhammer symbol. As a result, by substituting Eqs. \eqref{eq:aux1} and \eqref{eq:aux2} into Eq. \eqref{eq:I1tilde} we obtain
\begin{align}
\tilde I^{(1)}_{1} =& i \frac{\pi}{2} H^2 \sum_{j_1,\,j_2,\,j_3=0}^{\infty}\frac{2^{j_3 - 3 \nu} \,\csc(\nu \pi) \,\Gamma(1 + 2 \nu)}{\sqrt{\pi} \,\Gamma(1 + j_1) \,\Gamma(1 + j_3) \,\Gamma(1/2 + \nu) \,\Gamma(1 + \nu)} \nonumber \\
&  \times \Big(\frac{\left(\frac{k_1 + k_2 + k_3}{k_1}\right)^{1/2 - j_3 + \nu}\, \Gamma(1/2 + j_3 - \nu) \,\Gamma(7/2 + j_3 - \nu) \,\Gamma^2(1/2 + j_1 + \nu)}{\sqrt{2}\,(k_1 + k_2 + k_3)^4 \,\Gamma(1 + j_3 - 2 \nu) \, \Gamma(1 + j_1 + 2 \nu)} \nonumber\\
&  \qquad - \frac{2^{-1/2 +2 \nu} \,\left(\frac{k_1 + k_2 + k_3}{k_1}\right)^{1/2 - j_3 - \nu}\,
   \Gamma^2(1/2 + j_1 - \nu) \,\Gamma(1/2 + j_3 + \nu) \,\Gamma(7/2 + j_3 + \nu)}{(k_1 + k_2 + k_3)^4 \,\Gamma(1 + j_1 - 2 \nu)\,\Gamma(1 + j_3 + 2 \nu)}   \nonumber\\
&  \qquad - \frac{2^{j_1 + j_2 + \nu} \, \left(\frac{3 k_1 + k_2 + k_3}{k_1}\right)^{-j_1 - j_2 - j_3} \,\Gamma(4 + j_1 + j_2 + j_3) \,\Gamma(1/2 + j_3 - \nu)\,\Gamma^2(1/2 + j_1 +\nu)}{(3 k_1 + k_2 + k_3)^4 \,\Gamma(1 + j_3 - 2 \nu) \,\Gamma(3/2 + j_1 + j_2 + \nu)\,\Gamma(1 + j_1 + 2 \nu)}    \nonumber\\ 
&  \qquad+ \frac{2^{j_1 + j_2+\nu} \,\left(\frac{3 k_1 + k_2 + k_3}{k_1}\right)^{-j_1 - j_2 - j_3}\,\Gamma^2(1/2 + j_1 - \nu) \,\Gamma(1/2 + j_3 + \nu)\,\Gamma(4 + j_1 + j_2 + j_3)}{(3 k_1 + k_2 + k_3)^4 \,\Gamma(1 + j_1 - 2 \nu) \, \Gamma(3/2 + j_1 + j_2 - \nu)\, \Gamma(1 + j_3 + 2 \nu)}  \Big) \, . 
\end{align}
It follows that the nested integrals yield a purely imaginary contribution and, as such, do not affect the final result in Eq.~\eqref{eq:Tvar_GR1}. This outcome is not unexpected, as it aligns with recently established theorems on parity-odd correlators (see, e.g., \cite{Stefanyszyn:2023qov,Thavanesan:2025kyc}). In particular, it results from the fact that the mode functions are initialized in the Bunch-Davies vacuum and that the nested integrals are globally IR-convergent.

The factorized contribution can be calculated easily taking the limit $\tau_0 \rightarrow 0$ once we remind the following integral involving the Hankel functions (see e.g. \cite{Arkani-Hamed:2018kmz,Cabass:2022rhr})
\begin{align} 
I_{\nu,n}(k,q) &= \int_{-\infty}^{0}  d \tau \, (-\tau)^{n-1/2} \, H^{(2)}_\nu (-k \tau) \, e^{i q\tau} \nonumber \\
&= (-1)^{n+1} i^n \frac{2}{\sqrt{\pi}} e^{i (\frac{\pi}{2} \nu+\frac{\pi}{4})} \left(\frac{1}{2 k}\right)^{n+1/2} \, \frac{\Gamma(1/2+n-\nu) \Gamma(1/2+n+\nu)}{\Gamma(1+n)} \nonumber \\
& \qquad\qquad \times \,{\rm H2F1}\left(1/2+n-\nu, \,1/2+n+\nu, \, 1+n, \,\frac{1}{2} - \frac{q}{2 k}\right) \, , 
\end{align}
where ${\rm H2F1}\left(a, \,b, \, c, \,z\right)$ denotes the Gauss hypergeometric function. 

In the integral in Eq. \eqref{eq:I2tilde_decom} we have the case $n = 3$:
\begin{align} 
I_{\nu,3}(K_1^c,K_2^c+K_3^c) &= - i \frac{2}{\sqrt{\pi}} e^{i (\frac{\pi}{2} \nu+\frac{\pi}{4})} \left(\frac{1}{2 K_1^c}\right)^{7/2} \, \frac{\Gamma(7/2-\nu) \Gamma(7/2+\nu)}{\Gamma(4)} \nonumber \\
& \qquad\qquad \times \,{\rm H2F1}\left(7/2-\nu, \,7/2+\nu, \, 4, \,\frac{1}{2} - \frac{K_2^c+K_3^c}{2 K_1^c}\right) \, .
\end{align}
Using this result and performing the remaining integrals of the Bessel functions using the same strategy as specified above we get 
\begin{align} \label{eq:Itildefinalf}
&\tilde I^{(2)}_{1} = -\frac{H^2}{2} \cot{(\nu \pi)} \, \frac{1}{\left(2 K_1^c\right)^{4}} \, \frac{\Gamma(7/2-\nu) \Gamma(7/2+\nu)}{\Gamma(4)} \, \mathcal A(\nu) \nonumber \\
& \qquad\qquad \times \,{\rm H2F1}\left(7/2-\nu, \,7/2+\nu, \, 4, \,\frac{1}{2} - \frac{K_2^c+K_3^c}{2 K_1^c}\right)  \, ,
\end{align}
where 
\begin{align}
\mathcal A(\nu) = \sum_{j_1=0}^{\infty} \left( \frac{\Gamma^2(1/2+j_1+\nu)}{\Gamma(1+j_1)\Gamma(1+j_1+ 2\nu)} - \frac{\Gamma^2(1/2+j_1-\nu)}{\Gamma(1+j_1)\Gamma(1+j_1-2\nu)}\right) \, . 
\end{align}
$\mathcal A(\nu)$ is convergent and can be evaluated for each value of $\nu \neq 1/2$. We can relate this to the $\tan(\nu \pi)$ as 
\begin{align}
\mathcal A(\nu) = - 2 \pi \tan(\nu \pi) \, . 
\end{align}
So, in the end Eq. \eqref{eq:Itildefinalf} becomes
\begin{align} \label{eq:Itildefinal}
&\tilde I^{(2)}_{1} = H^2 \pi \, \frac{1}{\left(2 K_1^c\right)^{4}} \, \frac{\Gamma(7/2-\nu) \Gamma(7/2+\nu)}{\Gamma(4)} \, \,{\rm H2F1}\left(7/2-\nu, \,7/2+\nu, \, 4, \,\frac{1}{2} - \frac{K_2^c+K_3^c}{2 K_1^c}\right)  \, ,
\end{align}
which is well defined for each $0\leq \nu<3/2$. This quantity must be substituted in Eq. \eqref{eq:Tvar_GR1} in order to obtain the time-ordered variance in Eq. \eqref{eq:Tvar_GR1_computed}.

In evaluating the factorized contribution the IR-divergences for $\nu>1/2$ are given by the $j =0$ term of the series decomposition in Eq. \eqref{eq:Jseries} once we express the $Y$ function in terms of the $J$ function in the integral
\begin{align}
\Big[\int_{-\infty}^{\tau_0}  \,  d \tau_2 \,(-\tau_2)^{-1/2}\, Y_\nu(-K_1^c \tau_2)  \, e^{i K_1^c \tau_2}\Big] \, .
\end{align}
However, this term provides a real contribution once performing the time-integration, and therefore the IR-divergences are erased once we take the imaginary part of the integral as prescribed by Eq. \eqref{eq:I2tilde_decom}.  

Our computation is valid assuming that inflation lasts forever ($\tau_0 = 0$), allowing the spectator field of any non-zero mass to decay completely. However, if inflation does not last long enough (i.e. if $\tau_0$ is not small enough), one should add to the final result some additional terms proportional to $(-\tau_0)^{3/2-\nu}$ which account for the fact that the spectator field may not completely decay during inflation, for example when its mass is tiny (towards the limit $\nu \rightarrow 3/2$).

For the same reason, when $\nu = 3/2$ the way of performing the time-integrals which led to Eq. \eqref{eq:Itildefinal} is incorrect. Instead, in this case we need to employ the massless mode-function for the spectator field $\sigma$
\begin{align} \label{eq:mode_function_sigma}
v^{\sigma}_{\mathbf k}|_{\nu =3/2}= & \frac{H}{\sqrt{2 \, k^3}} (1+i k \tau) \, e^{-ik\tau} 
\end{align}
and then performing the time-integrations taking a finite duration time for inflation, $\tau_0\neq 0$. By substituting the mode-function \eqref{eq:mode_function_sigma} back in Eqs. \eqref{I1_comp} and  \eqref{I2_comp}, we can perform the time-integrations easily with the usual $i \epsilon$-prescription at infinity and we end up with a vanishing result, i.e.
\begin{align} 
I^{(1)}_{1}|_{\nu = 3/2}  + I^{(2)}_{1}|_{\nu = 3/2}  = 0 \, .
\end{align}

\section{Polarization tensor} \label{app:pol_ten}

By means of momentum conservation, $\mathbf{k_1}+\mathbf{k_1}+\mathbf{k_3}=0$, and invariance under rotations, we can choose the wave vectors appearing in the bispectrum in a way that they lie on the same plane. By imposing that they lie on the $(x,y)$ plane and that the graviton wave-vector $\mathbf{k_3}$ lies along the x-axis, we have
\begin{equation}
\mathbf{k}_1=k_1(\cos\bar\varphi_1,\sin\bar\varphi_1,0) \, ,\quad \mathbf{k}_2=k_2(\cos\bar\varphi_2,\sin\bar\varphi_2,0)\, ,\quad \mathbf{k}_3=k_3(1,0,0) \, ,
\end{equation}
where $\bar\varphi_1$ and $\bar\varphi_2$ are the angles that $\mathbf{k_3}$ forms with $\mathbf{k_1}$ and $\mathbf{k_2}$ respectively. We can express the angles in terms of triangle momenta as 
\begin{equation}
\cos\bar\varphi_2=\frac{k_1^2-k_2^2-k_3^2}{2 k_2 k_3}\, ,\quad \sin\bar\varphi_2=\frac{\xi}{2 k_2 k_3}\, ,\quad \cos\bar\varphi_1=\frac{k_2^2-k_3^2-k_1^2}{2 k_1 k_3}\, ,\quad \sin\bar\varphi_1=-\frac{\xi}{2 k_1 k_3}\, ,    
\end{equation}
with
\begin{equation}
\xi=\sqrt{2k_1^2k_2^2+2k_2^2k_3^2+2k_3^2k_1^2-k_1^4-k_2^4-k_3^4}\, .    
\end{equation}
Using the definitions \eqref{eq:helicity_tensor_pol} and \eqref{eq:linear_tensor_pol} we can then write the polarization tensor of the graviton of momentum $\mathbf{k_3}$ along the x-axis as 
\begin{equation}
e_{ij}^{\lambda_3}(\hat{k}_3)=\frac{1}{\sqrt{2}}
\begin{pmatrix}
0 & 0 & 0\\
0 & 1 & i \,\alpha_{\lambda_3}\\
0 & i \,\alpha_{\lambda_3} & -1\\
\end{pmatrix},
\end{equation}
where $\alpha_R= 1$ and $\alpha_L= - 1$, respectively. Also, when the field of momentum $\mathbf{k_2}$ is a graviton (in the computation of the $\langle\zeta h h \rangle$ bispectrum) its polarization tensor can be written as  
\begin{equation}
e_{ij}^{\lambda_2}(\hat{k}_2)=\frac{1}{\sqrt{2}}
\begin{pmatrix}
\sin^2\bar\varphi_2 & - \cos\bar\varphi_2\sin\bar\varphi_2 & - i \,\alpha_{\lambda_2}\sin\bar\varphi_2\\
- \cos\bar\varphi_2\sin\bar\varphi_2 & \cos^2\bar\varphi_2 & i \,\alpha_{\lambda_2} \cos\bar\varphi_2\\
-i \,\alpha_{\lambda_2}\sin\bar\varphi_2  & i \,\alpha_{\lambda_2} \cos\bar\varphi_2 & -1\\
\end{pmatrix} \, .
\end{equation}
It follows that the Latin contractions in Eqs. \eqref{eq:sst_bisp} and \eqref{eq:stt_bisp} read 
\begin{align}
e^{\lambda}_{ij}(\hat k_3) \, k_2^i \, k_2^j  = \frac{\xi^2}{4 \sqrt 2 \, k^2_3} \, , 
\end{align}
\begin{align}
e^{\lambda_2}_{ij}(\hat k_2)e^{\lambda_3}_{ij}(\hat k_3) = \begin{cases} \frac{1}{2} \left(1 - \cos(\bar\varphi_2)\right)^2 \quad \mbox{when } \lambda_2 = \lambda_3 \\
\frac{1}{2} \left(1 + \cos(\bar\varphi_2)\right)^2  \quad\mbox{when } \lambda_2 \neq \lambda_3
\end{cases}\, , 
\end{align}
and
\begin{align}
k_{2}^l \, k_{3}^m \,e^{\lambda_2}_{im}(\hat k_2)e^{\lambda_3}_{il}(\hat k_3) = \begin{cases} \frac{1}{2} k_2 k_3 \,\sin^2(\bar\varphi_2) \, (1-\cos(\bar\varphi_2)) \quad \mbox{when } \lambda_2 = \lambda_3 \\
-\frac{1}{2} k_2 k_3 \,\sin^2(\bar\varphi_2) \, (1+\cos(\bar\varphi_2)) \quad \mbox{when } \lambda_2 \neq \lambda_3
\end{cases}\, .
\end{align}
In terms of the raw momenta these can be written as 
\begin{align}
e^{\lambda}_{ij}(\hat k_3) \, k_2^i \, k_2^j  = \frac{2k_1^2k_2^2+2k_2^2k_3^2+2k_3^2k_1^2-k_1^4-k_2^4-k_3^4}{4 \sqrt 2 \, k^2_3} \, , 
\end{align}
\begin{align}
e^{\lambda_2}_{ij}(\hat k_2)e^{\lambda_3}_{ij}(\hat k_3) =  \begin{cases} \frac{(k_1^2 -( k_2 + k_3)^2)^2}{8 k_2^2 k_3^2} \quad \mbox{when } \lambda_2 = \lambda_3 \\
\frac{(k_1^2 -( k_2 - k_3)^2)^2}{8 k_2^2 k_3^2} \quad\mbox{when } \lambda_2 \neq \lambda_3
\end{cases}\, , 
\end{align}
and
\begin{align}
k_{2}^l \, k_{3}^m \,e^{\lambda_2}_{im}(\hat k_2)e^{\lambda_3}_{il}(\hat k_3) =  \begin{cases} \frac{(k_1 + k_2 - k_3) (k_1 - k_2 +k_3) (-k_1 + k_2 + k_3)^2 (k_1 + k_2 + k_3)^2}{16 k_2^2 k_3^2} \quad  \mbox{when } \lambda_2 = \lambda_3 \\
-\frac{(k_1^2 - (k_2 - k_3)^2)^2 (-k_1^2 + (k_2 + k_3)^2)}{16 k_2^2 k_3^2} \quad \mbox{when } \quad \lambda_2 \neq \lambda_3
\end{cases}\, .
\end{align}

\bibliographystyle{hunsrt}
\bibliography{CS_iso}

\end{document}